\documentstyle[12pt,epsf]{article}
\input psfig.sty

\begin{document}
\begin{titlepage}
\begin{flushright}
{\tt hep-th/0005166}\\
{\tt  TIFR/TH/00-22}~~ \\
May, 2000 
\end{flushright}
\vfill
\begin{center}
{\Large \bf D-brane gauge theories from toric singularities
of the form ${\bf C^3}/\Gamma$ and ${\bf C^4}/\Gamma$} \\[1cm]
Tapobrata Sarkar\footnote{Email: tapo@theory.tifr.res.in}\\
{\em Department of Theoretical Physics, \\ Tata Institute of 
Fundamental Research, \\ Homi Bhabha Road, Mumbai 400 005, India}\\ 
\end{center}
\vspace{2cm}
\begin{abstract}

We discuss examples of D-branes probing toric singularities, 
and the computation of their world-volume gauge theories from the
geometric data of the singularities. We consider several such 
examples of D-branes on partial resolutions of the orbifolds 
${\bf C^3/Z_2\times Z_2}$,${\bf C^3/Z_2\times Z_3}$ and 
${\bf C^4/Z_2\times Z_2 \times Z_2}$.

\end{abstract}
\vfill
\end{titlepage}
\section{Introduction}

In the last few years, a lot of progress has been made 
in the study of D-branes
as probes of sub-stringy geometry. While the usual picture of
space-time is supposed to be valid upto the string scale, it
was argued in \cite{dkps} that D-branes are the natural candidates
to probe the geometry of space-time beyond the string scale. 
As has been widely accepted by now, the standard concepts of
space-time appears, from the D-brane perspective, as the vacuum
moduli space of its world volume gauge theory.
Indeed, it has been found that the space-time appearing in D-brane 
probes at ultra-short distance scales has qualitatively different 
features from that probed by bulk strings \cite{gk}. 

Of particular interest has been the study of D-branes on Calabi-Yau
manifolds. In \cite{agm},\cite{wittenphases}, the
moduli space of Calabi-Yau manifolds was studied using  
fundamental strings. A rich underlying geometric structure was discovered,
and the moduli space was shown to contain topologically distinct
geometric (Calabi-Yau) phases, as well as non-geometric 
(Landau-Ginzburg) phases. The K\"ahler sector of the vacuum moduli space
was shown to consist of several domains (separated by singular 
Calabi-Yau spaces), whose union is isomorphic to the complex structure
moduli space of the mirror manifold. It was also shown, that using
mirror symmetry as in \cite{agm} or the approach of gauged linear sigma
models as in \cite{wittenphases}, one can interpolate smoothly between
the topologically distinct points in the K\"ahler moduli space. 
In the fundamental string picture, the geometric and non-geometric
phases of Calabi-Yau manifolds appear as two distinct possibilities, 
with the latter being thought of as an analytic continuation of the 
geometric phases in the presence of a non-zero theta angle in the 
language of \cite{wittenphases}. It was, however, shown in \cite{dgm}
that the presence of additional open string sectors change this 
picture considerably, and that the D-brane linear sigma model probes
only a part of the full linear sigma model vacuum moduli space, namely
the geometric phase. It was explicitly demonstrated in \cite{dgm} that
D-branes (in the particular examples of the orbifold ${\bf C^3/Z_3}$
and ${\bf C^3/Z_5}$) `project out' the non-geometric phase, in agreement with 
the results obtained in \cite{wittenm}. This was found to be 
generally true for Calabi-Yau orbifolds of the form ${\bf C^3/Z_n}$ 
\cite{muto}.

Since the pioneering work of \cite{dm}, which dealt with D-branes on
Abelian orbifold singularities of the form ${\bf C^2/Z_n}$ that was
generalised to arbitrary non-Abelian singularities of the form
${\bf C^2}/\Gamma$ ($\Gamma$ being a subgroup of $SU(2)$),  
D-branes on Abelian and non-Abelian orbifold 
backgrounds have been extensively studied in \cite{muto},\cite{jm},
\cite{dg1},\cite{dko},\cite{muto1},\cite{fhh1}. D-branes at
orbifolded conifold singularities have been considered 
in \cite{ot}. (See also \cite{celw} for an analysis of a blowup of 
the four-dimensional $N=1$, ${\bf Z_3}$ orientifold). 
Of particular interest has been the applications of methods of toric
geometry \cite{fulton},\cite{cox} to the study of D-brane gauge theories
on such singularities. In the approach pioneered in \cite{dgm},
the matter content and the interactions of the D-brane gauge theory, 
which are specified by the 
D-terms and the F-terms of the gauge theory respectively, are treated 
on the same footing, and the gauge theory information can be encoded
as algebraic equations of the toric variety. 
It is an interesting question to ask, if this procedure can be
reversed, i.e, given a particular toric singularity, is there a way
of consistently reading off the world volume gauge theory data for a D-brane
that probes this singularity. A step in this direction has been recently 
taken in \cite{fhh}. The authors of \cite{fhh} have given an algorithm
by which the geometrical data encoded in a toric diagram can be used
to construct the matter content and superpotential interaction of the
D-brane gauge theory that probes it, and have demonstrated this
in the cases of the suspended pinch point (SPP) singularity of the 
blowup of the ${\bf C^3/Z_2\times Z_2}$ orbifold and the various 
partial resolutions of the ${\bf C^3/Z_3\times Z_3}$ orbifold 
analysed in \cite{bglp}. In the present paper, we critically examine 
this procedure in several other examples. We first consider this 
algorithm, which we call the
inverse toric procedure, for partial resolutions of the orbifold 
${\bf C^3/Z_2\times Z_2}$. Further, we consider several blowups of the 
orbifolds ${\bf C^3/Z_2\times Z_3}$ and ${\bf C^4/Z_2\times Z_2\times 
Z_2}$ and present results on the extraction of the D-brane gauge theory 
data from the given toric geometric data.

The organisation of the paper is as follows. In section 2, we first 
recapitulate the essential details of the construction of the D-brane 
gauge theory on partial resolutions of the orbifold ${\bf C^3/Z_2\times 
Z_2}$ \cite{gr1},\cite{mr} that will also set the notation and 
conventions to be used in the rest
of the paper and then demonstrate the inverse toric algorithm 
to construct the gauge theory of a D-brane probing the ${\bf Z_2}$
and the conifold singularities obtained by partially resolving 
the  ${\bf C^3/Z_2\times Z_2}$ orbifold. 
Section 3 deals with the orbifold
${\bf C^3/Z_2\times Z_3}$ and its various partial resolutions. 
In section 4, we consider the example of the orbifold 
${\bf C^4/Z_2\times Z_2\times Z_2}$ 
and consider some resolutions of the same. Section 5 contains 
discussions and directions for future work.

\section{D-branes on ${\bf C^3/Z_2\times Z_2}$ and its blowups} 
 
We begin this section by recapitulating the essential details of 
the application of toric methods to the construction of D3-branes
on ${\bf C^3/Z_2\times Z_2}$ and some of its partial resolutions 
which will also set the notations and conventions to be followed in
the rest of the paper. 
Recall that the world volume supersymmetric gauge theory of D-branes 
probing a toric singularity are characterised by its matter content 
and interactions. While the former are given by the
D-term equations, the latter are specified, via the superpotential,
as the F-term equations. These two sets of equations, in conjunction,
define the moduli space of the theory. In order to define
the D-brane moduli space by toric methods, the standard procedure
\cite{dgm} is to re-express the F and D-term constraints in terms
of a set of homogeneous variables $p_{\alpha}$, and a concatenation
of these two sets of equations give rise to the toric data. 
Let us briefly review this construction for the example of D-3
branes on the orbifold ${\bf C^3/(Z_2\times Z_2)}$.

We start by considering the field theory of a set of D-3 branes
at a ${\bf C^3/(Z_2\times Z_2)}$ singularity. Generically, a single
D-p brane of type II string theory at a point in the orbifold
${\bf C^3}/\Gamma$ is constructed by considering the theory of $|\Gamma|$
D-p branes in ${\bf C^3}$ and then projecting to ${\bf C^3}/\Gamma$
where the projection is defined by a combined action of 
$\Gamma$ on ${\bf C^3}$ as well as the D-brane Chan-Paton index.
The unbroken supersymmetry in $d=4$ would then be $N=2$ in the closed
string sector while the open string sector will have $N=1$ supersymmetry.
In our example, a generator $g$ of the discrete 
group ${\bf Z_2\times Z_2}$ acts simultaneously on the coordinates
of ${\bf C^3}$ and also the D-3 brane 
Chan-Paton factors. The surviving fields in the theory are the ones
that are left invariant by a combination of these two actions. 
Specifically, denoting the action of the quotienting group on ${\bf C^3}$
by $R(g)$ and that on the Chan-Paton indices by $\gamma(g)$ (where
$\gamma$ denotes a regular representation of $g$), 
the surviving components of the scalars $X$ that live on the brane
world volume are those for which $X=R(g)\gamma(g)X\gamma(g)^{-1}$, and the 
components of the gauge fields $A$ that survive the projection are those 
that satisfy $A=\gamma(g)A\gamma(g)^{-1}$. After imposing these 
projections, the $N=1$ matter multiplet of the theory consists of the 
twelve fields
\begin{equation}
\left(X_{13},X_{31},X_{24},X_{42}\right),~
\left(Y_{12},Y_{21},Y_{34},Y_{43}\right),~ 
\left(Z_{14},Z_{41},Z_{23},Z_{32}\right)
\label{nonzerox}
\end{equation}

Where we have denoted the complex bosonic fields corresponding
to the position coordinates tangential to ${\bf C^3}$ by $X^1=X,
X^2=Y,X^3=Z$.  After projection of the gauge field, the unbroken 
gauge symmetry in this case is the subgroup $U(1)^4$ of $U(4)$, and, 
apart from an overall $U(1)$, we have a $U(1)^3$ theory. The 
total charge matrix is given by

{\small
\begin{eqnarray}
d=\pmatrix{
-1&0&0&1&-1&1&0&0&-1&0&1&0\cr
0&-1&1&0&1&-1&0&0&0&-1&0&1\cr
0&1&-1&0&0&0&-1&1&1&0&-1&0\cr
1&0&0&-1&0&0&1&-1&0&1&0&-1\cr
}\end{eqnarray}}

With the last row of $d$ signifying the overall $U(1)$. 
The D-term constraints are now given by
\begin{equation}
\sum_{\mu} d_{(i)\mu}^\beta|X^{\mu}_{\beta}|^2 - \zeta_i=0
\end{equation}
where $d_{(i)\mu}^\beta$ is the charge of $X^{\mu}_{\beta}$ under
the $i$'th $U(1)$ and $\beta$ signifies the matrix
indices of the surviving components of $X^{\mu}$. $\sum_i\zeta_i=0$ 
is the condition for the existence of supersymmetric vacua. 

The superpotential of the theory is given by
the expression
\begin{equation}
W={\mbox{Tr}}\left[X^1,X^2\right]X^3
\end{equation}
With the vacuum satisfying 
\begin{equation}
\frac{\partial W}{\partial X^{\mu}}=0;~~~{\mbox{i.e}}~~~
\left[X^{\mu},X^{\nu}\right]=0
\label{superpot}
\end{equation}

Using the expressions for the $X^{\mu}$'s from eq.(\ref{nonzerox})
in eq. (\ref{superpot}), one can derive twelve constraints on the 
twelve fields $X_{ij},Y_{ij},Z_{ij}$, out of which six are seen to 
be independent, 
hence the twelve initial fields can be solved in terms of 
six independent fields.
Denoting the twelve matter fields collectively by $X_i, i=1\cdots 12$, 
and the set of six independent fields (which we take to be
$X_{13},X_{24},Y_{21},Y_{34},Z_{14},Z_{32}$) by $x_b, (b=1\cdots 6)$, 
the solution, which is of the form 
\begin{equation}
X_i=\prod_{b=1}^6 x_b^{K_{ib}}
\label{indep}
\end{equation}
can be encoded in the columns of the matrix $K$, given by 
{\small
\begin{eqnarray}
K=\pmatrix{
~&X_{13}&X_{24}&X_{31}&X_{42}&Y_{12}&Y_{21}&Y_{34}&Y_{43}&Z_{14}&
Z_{23}&Z_{32}&Z_{41}\cr
X_{13}&1&0&0&1&1&0&0&1&0&0&0&0\cr
X_{24}&0&1&1&0&-1&0&0&-1&0&0&0&0\cr
Y_{21}&0&0&0&0&0&1&0&1&0&1&0&1\cr
Y_{34}&0&0&0&0&1&0&1&0&0&-1&0&-1\cr
Z_{14}&0&0&-1&-1&0&0&0&0&1&1&0&0\cr
Z_{32}&0&0&1&1&0&0&0&0&0&0&1&1\cr
}\end{eqnarray}}

The columns of the above matrix are vectors in the lattice 
$N={\bf Z^6}$, and define the edges of a cone $\sigma$. In order to make 
the description of the toric variety explicit, we consider the dual cone 
$\sigma^{\vee}$, which is defined in the dual lattice $M={\mbox{Hom}}
(N,{\bf Z})$.  The dual cone $\sigma^{\vee}$ is defined to be the set of 
vectors in the dual lattice $M$ which have non-negative inner products 
with the vectors of $\sigma$, i.e
\begin{equation}
\sigma^{\vee}=\{{\bf m} \in M: \langle{\bf m},{\bf n}\rangle
 \geq 0~\forall {\bf n} \in \sigma\}
\end{equation}

The dual cone corresponds to defining a new set of fields
$p_{\alpha} (\alpha=1\cdots c)$, in terms of which the variables $x_b$ 
are solved, in a way that is consistent with the relation (\ref{indep}). 
In this example, $c=9$ and the dual cone is generated by the columns of
the matrix 
{\small
\begin{eqnarray}
T=\pmatrix{
~&p_1&p_2&p_3&p_4&p_5&p_6&p_7&p_8&p_9\cr
X_{13}&1&1&0&0&0&0&0&0&1\cr
X_{24}&0&1&1&0&0&0&0&0&1\cr
Y_{21}&0&0&1&1&1&0&0&0&0\cr
Y_{34}&0&0&1&0&1&1&0&0&0\cr
Z_{14}&0&0&0&0&0&1&1&0&1\cr
Z_{32}&0&0&0&0&0&1&1&1&0\cr
}\end{eqnarray}}

The relationship between $p_{\alpha}$ and $x_b$ are
then of the form
\begin{equation}
x_b=\prod_{\alpha}p_{\alpha}^{T_{b\alpha}}
\end{equation}
Thus, there are nine variables $p_{\alpha}$
in terms of which we parametrize the six independent variables 
$x_b$.  This parametrization, however, has redundancies, and in order
to eliminate these, we introduce a new set of ${\bf C^*}$ actions on the
$p_{\alpha}$'s, with the condition that under these actions, the
fields $x_b$ are left invariant. This can be 
equivalently described by the introduction of a new set of $U(1)$ gauge
symmetries, such that the $x_b$'s are gauge invariant \cite{agm}. 
In our example, since there are nine fields $p_{\alpha}$ and six 
independent variables $x_b$, we introduce a gauge group $U(1)^3$, such 
that the $p_{\alpha}$'s are charged with respect to these, with 
charges $Q_{n\alpha}$ for the $n$th $U(1)$. The gauge invariance 
condition demands that the charge matrix $Q$ must obey
\begin{equation}
TQ^t=0
\end{equation}
where $Q^t$ is the transpose of $Q$. From the matrix $T$ given 
above, the charge matrix is given by
{\small
\begin{eqnarray}
Q=\pmatrix{
~&p_1&p_2&p_3&p_4&p_5&p_6&p_7&p_8&p_9\cr
U(1)_1&0&0&0&1&-1&1&-1&0&0\cr
U(1)_2&0&1&0&0&0&0&1&-1&-1\cr
U(1)_3&1&-1&1&0&-1&0&0&0&0\cr
}\end{eqnarray}}

We now consider the constraints imposed on the original fields
$X_i$ via the D-flatness conditions, and impose these conditions in
terms of the new fields $p_{\alpha}$. The charges of the 
fields $x_b$ under the original $U(1)$ charge assignments are given 
by the matrix

{\small
\begin{eqnarray}
V=\pmatrix{
x_1&x_2&x_3&x_4&x_5&x_6\cr
1&0&-1&0&1&0\cr
0&1&1&0&0&-1\cr
-1&0&0&1&0&1\cr
}\label{vmatrix}\end{eqnarray}}
From the relationship $x_b=\prod_{\alpha}p_{\alpha}^{T_{b\alpha}}$, it is
clear that in order for the charge assignment $q^p_{i\alpha}$ for $p_{\alpha}$ 
(corresponding to the original set of $(i=3)$ $U(1)$ symmetries) to reproduce
that for the $x_b$'s, we must have $q^pT^t=V$, and one possible choice
for the matrix $q^p$ is  $q^p_{i\alpha}= (VU)_{i\alpha}$ with the matrix 
$U$ satisfying $TU^t={\mbox{Id}}$.  
The matrix $U$ is thus the inverse of the matrix $T^t$. However,
since $T^t$ is a rectangular matrix the usual definition of its inverse
(as in the case of a square matrix) 
is not applicable and there are several choices for $U$. In particular, 
for a rectangular matrix $A$, using the rules of single value
decomposition, we can compute the Moore-Penrose inverse of $A$
which is defined in a way that the sum of the squares of all the entries 
in the matrix $(AA^{-1}-{\mbox{Id}})$ is minimised, 
where ${\mbox{Id}}$ is the identity matrix in appropriate 
dimensions. For any choice of the inverse of $T^t$, the final toric data 
will of course be identical, and in this section and the next, we 
simply use the results of \cite{pru} in defining the
matrix $U$. For this case, the matrix $U$ is 

{\small
\begin{eqnarray}
U=(T^t)^{-1}=\pmatrix{
1&0&0&0&0&0&0&0&0\cr 
-1&1&0&0&0&0&0&0&0\cr
0&0&0&1&0&0&0&0&0 \cr
0&0&0&0&0&1&-1&0&0\cr
0&-1&0&0&0&0&0&0&1 \cr
0&0&0&0&0&0&0&1&0\cr
}
\nonumber
\end{eqnarray}
}

\begin{figure}
\centering
\epsfxsize=3.5in
\hspace*{0in}\vspace*{.2in}
\epsffile{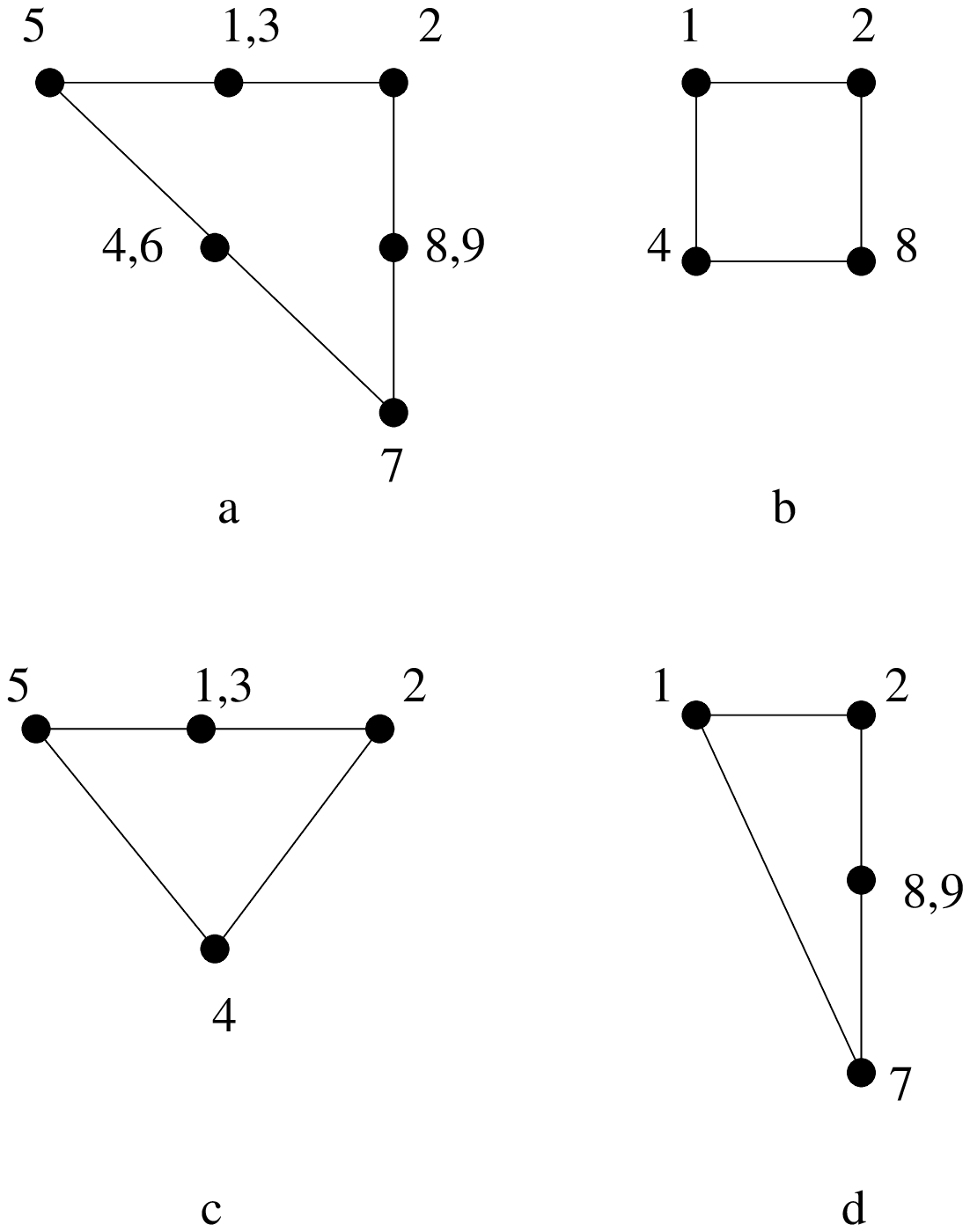}
\caption{\small Toric diagram showing a) the 
${\bf C^3/Z_2\times Z_2}$ orbifold, b) a resolution to the conifold,
c) resolution to the ${\bf Z_2}$ orbifold singularity and d) a second
resolution to the ${\bf Z_2}$ orbifold singularity. We have also marked
the fields that remain dynamical in these resolutions.} 
\label{322figs}
\end{figure}

Now, multiplying by $V$ and concatenating $Q$ and $VU$ we obtain the 
total charge matrix,
{\small
\begin{eqnarray}
Q_t=\pmatrix{
0&0&0&1&-1&1&-1&0&0&0\cr
0&1&0&0&0&0&1&-1&-1&0\cr
1&-1&1&0&-1&0&0&0&0&0\cr
1&-1&0&-1&0&0&0&0&1&\zeta_1\cr   
-1&1&0&1&0&0&0&-1&0&\zeta_2\cr   
-1&0&0&0&0&1&-1&1&0&\zeta_3\cr
}\label{fullq}\end{eqnarray}
}

where the $\zeta_i$ are the Fayet-Illiapoulos parameters for the 
original $U(1)$'s, there being three independent $\zeta_i$'s because 
of the constraint $\sum_{i=1}^4\zeta_i=0$. The toric data can now be 
calculated from the cokernel of the transpose of $Q_t$, which (after a 
few row operations) become
{\small
\begin{eqnarray}
{\tilde T}=\pmatrix{
0&1&0&0&-1&0&1&1&1\cr
1&1&1&0&1&0&-1&0&1\cr
1&1&1&1&1&1&1&1&1\cr
}\label{fulltoricdata}\end{eqnarray}}
Now from the dual cone $T$, we can, via the matrix $K$, 
find expressions for all the twelve initial fields in terms 
of the fields $p_{\alpha}$. 
These are then used to define invariant variables in the following way

\begin{eqnarray}
X_{13}X_{31}=x&=&p_1p_2^2p_3p_8p_9\nonumber\\
Y_{12}Y_{21}=y&=&p_1p_3p_4p_5^2p_6\nonumber\\
Z_{14}Z_{41}=z&=&p_4p_6p_7^2p_8p_9\nonumber\\
X_{13}Y_{34}Z_{41}=w&=&p_1p_2p_3p_4p_5p_6p_7p_8p_9
\end{eqnarray}

which follow the relation $xyz=w^2$. The toric diagram of the orbifold
${\bf C^3/Z_2\times Z_2}$ is shown in figure \ref{322figs}(a). 
Let us now consider some examples of the possible partial
resolutions of this space. We will discuss three possible partial blowups 
as in \cite{pru},\cite{mp}.

The first example corresponds to blowing up the orbifold 
${\bf C^3/Z_2\times Z_2}$
by a ${\bf P}_1$ parametrised by $w'=\frac{w}{z}$. This gives the suspended
pinch point (SPP) singularity given by $xy=zw'^2$. 
The SPP can be further blown up by introducing a second ${\bf P}_1$.
One possibility is to parametrise this ${\bf P}_1$ by $y'=\frac{y}
{w'}$ (or $x'=\frac{x}{w'}$) and the remaining singularity is the 
conifold $x'y=zw'$ (or $xy'=zw'$) of figure \ref{322figs}(b). One can
also introduce a  ${\bf P}_1$ parametrised by say $x'=\frac{x}{z}$
whence the resulting blowup is a ${\bf C^2/Z_2\times C}$. Two possible 
toric diagrams of the latter are shown in fig. \ref{322figs}(c) and  
fig. \ref{322figs}(d).

Let us now come to the description of the inverse toric problem. Suppose
we are given with a singularity depicted by a toric diagram of the
form given in fig. \ref{322figs}(b), \ref{322figs}(c) or 
\ref{322figs}(d). We wish to extract the gauge theory 
data of a D-brane that probes such a singularity, starting from 
figure \ref{322figs}(a). This issue was 
addressed in \cite{fhh}, and let us review the basic steps in their 
algorithm. 

Recall that the toric data for a given singularity that a D-brane probes
is given by the matrix ${\tilde T}$ which 
is the transpose of the kernel of the total charge matrix (obtained by
concatenating the F and D-term constraints expressed in terms of the
homogeneous coordinates $p_{\alpha}$). Hence, the solution
to the inverse problem would imply the 
construction of an appropriately reduced total charge matrix 
$Q_t^{red}$, the cokernel of the transpose of which is 
the reduced matrix ${\tilde T_{red}}$ obtained by deleting the 
appropriate  columns (corresponding to the fields that have been 
resolved) from ${\tilde T}$. 
However, although ${\tilde T_{red}}$ is known, there is no unique 
way of specifying the reduced charge matrix $Q_t^{red}$ from a knowledge
of ${\tilde T_{red}}$. Further, there is no guarantee apriori that 
such a $Q_t^{red}$, if obtained, would continue to describe the world
volume theory of a D-brane that probes the given singularity.  The 
algorithm of \cite{fhh} gives a canonical method by which these issues
can be addressed. It consists of determining the fields to be resolved
by tuning the Fayet-Illiopoulos parameters appropriately (so that
the resulting theory at the end is still a physical D-brane gauge
theory) and then reading off the the matrix $Q_t^{red}$ which can 
be separated into the F and D-terms and thus can be used to obtain
the matter content and the superpotential determining the gauge theory.
Let us discuss this in some more details. The toric diagrams for the
partially resolved singularities are obtained by the deletion of 
of a subset of nodes from the original one. Conversely, a given toric
diagram of dimension $k$ can be embedded into a singularity of the
form ${\bf C}^k/\Gamma(k,n)$ where $\Gamma(k,n)={\bf Z_n\times Z_n\times} 
\cdots (k-1~{\mbox{times}})\cdots {\bf Z_n}$. In our case, the 
singularities mentioned above can be embedded into the toric
diagram for the singularity ${\bf C^3}/\Gamma(3,2)$. For example, the 
SPP, and the conifold singularity can both be embedded
into the ${\bf C^3/Z_2\times Z_2}$ singularity, as is obvious from their
toric diagrams. We shall henceforth refer to these singularities, in 
which we embed the partially resolved ones as the parent singularities.
A toric diagram corresponding to a partial resolution can of course
be embedded in more than one parent singularity and according to the
algorithm of \cite{fhh}, one can choose the minimal embedding. We
will return to this issue in a while.

We now determine the fields to be resolved from the parent 
singularity in order to reach the given toric diagram, 
by appropriately choosing the values of $\zeta_i$. The difficulty
that arises here is that given the toric diagram of the singularity
that we are interested in, there is no unique way of knowing exactly 
which fields (in the toric data of the parent singularity) have to
be resolved in order to reach the toric singularity of interest. 
Given a toric diagram for the parent singularity, a particular
resolution, along with the elimination of one or
more nodes, might also require a subset of fields from some other 
nodes to be resolved. Hence, we have to carefully tune the
FI parameters for a consistent blowup of the parent singularity to the
one that we are interested in. We now illustrate this procedure 
by the example of the blowup of the ${\bf C^3/Z_2\times Z_2}$ singularity 
to the  ${\bf C^2/Z_2\times C}$ orbifold illustrated in fig. \ref{322figs}(c). 
This example is, in a sense, straightforward, because from the toric 
diagram of fig. \ref{322figs}(c), it is clear that this being the ${\bf Z_2}$
singularity, we can directly read off the matter content and the interactions
from the standard results for D3-branes on the ${\bf Z_2}$ orbifold \cite{dm}. 
However, note that this is the simplest example where the inverse 
procedure of calculating the D-brane gauge theory has to be valid. 
The ${\bf Z_2}$ orbifold can be thought of as partial resolutions of the 
orbifolds ${\bf C^3/Z_2\times Z_2}$,  
or ${\bf C^3/Z_2\times Z_3}$, and the inverse toric procedure must give 
familiar results for both in agreement with \cite{dm},\cite{pru}. 
A lower dimensional toric orbifold can always be embedded in the toric
diagram of one of higher dimension, and the inverse toric procedure must
consistently reproduce the D-brane gauge theory on the former, starting 
from the latter data. As we will see, this is indeed the case. 

We proceed by embedding the ${\bf Z_2}$ singularity in 
${\bf C^3/Z_2\times Z_2}$. 
In order to determine the fields to be resolved, we perform 
Gaussian row reduction on the the matrix $Q_t$ to obtain 
{\small
\begin{eqnarray}
\pmatrix{
-1& 0& 0& 0& 0& 1& -1& 0& 1& \zeta_1 + \zeta_2 + \zeta_3\cr
0& -1& 0& 0& 0& 0& -1& 0& 2& \zeta_1 + \zeta_2\cr
0& 0& -1& 0& 0& 1& -1& 0& 1& \zeta_1\cr
0& 0& 0& -1& 0& 1& 0& 0& 0& \zeta_1 + \zeta_3\cr
0& 0& 0& 0& -1& 2& -1& 0& 0& \zeta_1 + \zeta_3\cr
0& 0& 0& 0& 0& 0& 0& -1& 1& \zeta_1 + \zeta_2\cr
}\nonumber\end{eqnarray}}

The above set of equations imply the constraints 
\begin{eqnarray}
-x_1+x_6-x_7+x_9&=&(\zeta_1+\zeta_2+\zeta_3)\nonumber\\
-x_2-x_7+2x_9&=&(\zeta_1+\zeta_2)\nonumber\\
-x_3+x_6-x_7+x_9&=&\zeta_1\nonumber\\
-x_4+x_6&=&(\zeta_1+\zeta_3)\nonumber\\
-x_5+2x_6-x_7&=&(\zeta_1+\zeta_3)\nonumber\\
-x_8+x_9&=&(\zeta_1+\zeta_2)
\end{eqnarray}

Where, for convenience of notation,  we have labelled $p_i^2=x_i$.
This set of equations is now solved in terms of the fields that we know
would definitely get resolved. From the toric diagram, three such
fields are $p_7,p_8,p_9$. However, from the relation between $x_8$ and
$x_9$ above, it suffices to solve the above set of equations in terms of
$x_7$ and $x_9$, say. This gives the solution set,
\begin{eqnarray}
\left[x_1,\cdots x_9\right]&=&\left[x_1,(2x_9-x_7-\zeta_1-\zeta_2),(x_1
+\zeta_2+\zeta_3),
(x_1+x_7-x_9+\zeta_2),\right.\nonumber\\
&~&(2x_1+x_7-2x_9+\zeta_1+2\zeta_2+\zeta_3)
(x_1+x_7-x_9+\zeta_1+\zeta_2+\zeta_3),\nonumber\\
&~&\left.x_7,(x_9-
\zeta_1-\zeta_2), x_9\right]
\end{eqnarray}
Now, from the toric diagram of fig.\ref{322figs}(c), we choose the
fields $[p_1,p_2,p_3,p_5]$ to have zero vev i.e these fields continue
to remain dynamical. 
Thus, in the above solution set, we impose $x_1,x_2,x_3,x_5=0$ 
to obtain
\begin{equation}
\left[x_1,\cdots x_9\right]=\left[0,0,0,(x_9-\zeta_1),0,(x_7-x_9+\zeta_1),x_7,
(x_9-\zeta_1-\zeta_2),x_9\right]
\end{equation}

We further set $x_9=\zeta_1$ in order to make $x_4=0$. Hence,
$x_6=x_7=\zeta_1-\zeta_2$. Also, $x_8=-\zeta_2$. Therefore, the fields 
to be resolved are $[p_6,p_7,p_8,p_9]$, all of which have positive
vevs. 

We now need to obtain the reduced charge matrix for this choice of 
variables to be resolved. The method of doing this essentially consists of
performing row operations on the full charge matrix $Q_t$ in eq. 
(\ref{fullq}), such that the 
columns corresponding to the resolved fields $[p_6,p_7,p_8,p_9]$ 
are set to zero. The reduced
charge matrix $Q_t^{red}$ thus obtained must be in the nullspace of the 
reduced toric matrix ${\tilde T}_{red}$, that can be directly evaluated 
from the full toric matrix in eq. (\ref{fulltoricdata}) by removing the 
columns $6,7,8,9$. Let us first discuss the general formalism for 
achieving this, which can then be easily applied to our present
example. Consider a general parent theory that has a total charge matrix 
$Q_t$ of dimensions $(a\times b)$, and suppose we wish to eliminate
$d$ fields from the parent theory in order to reach the theory of 
interest. This implies that we have to perform row operations on 
$Q_t$ so that the given $d$ columns will have zero entries and thus
be eliminated. The latter can be achieved by constructing the nullspace
of the transpose of a submatrix that is formed by precisely the $d$ columns
that need to be eliminated. Further, since the toric data ${\tilde T}$
was initially in the nullspace of $Q_t$, removal of the $d$ columns
from $Q_t$ to obtain $Q_t^{red}$ would mean that the reduced
toric data ${\tilde T_{red}}$ with the same $d$ columns removed would
be in the nullspace of $Q_t^{red}$. Thus, the expression
for $Q_t^{red}$ is given by \cite{fhh} 
\begin{equation}
Q^{red}_t=\left[{\mbox{NullSpace}}(Q_d^t)\cdot Q_r\right]
\label{qreduced}
\end{equation}
where $Q_d^t$ is the transpose of the matrix obtained from $Q$ 
containing as its columns, the deleted columns of $Q$ and $Q_r$
is the remaining matrix. 

In our example, we evaluate $Q_{red}$ directly,
by writing $Q_d^t$ as the transpose of the matrix that contains 
columns $6,7,8,9$ of the charge matrix $Q_t$, and $Q_r$ being
the matrix containing the columns $1,2,3,4,5,6,10$ of $Q_t$. This 
implies that the reduced charge matrix is
{\small
\begin{eqnarray}
Q_t^{red}=\pmatrix{
1& -1& 1& 0& -1& 0\cr
-2& 1& 0& 0& 1& \zeta_2 + \zeta_3\cr
}\end{eqnarray} }
From this, the reduced charge matrices corresponding to the
F and D-terms are 
\begin{equation}
Q_{red}=\left(1~~-1~~1~~0~~-1\right);~~~~(VU)_{red}
=\left(-2~~1~~0~~0~~1\right)
\end{equation}
Denoting the kernel of the charge matrix by $T_{red}$, the transpose
of its dual is the reduced $K^t$ matrix, 
{\small
\begin{eqnarray}
K_{red}^t=\pmatrix{
0& 1& 0& 1& 0\cr
1& 0& 0& 0& 0\cr
0& 1& 1& 0& 0\cr
0& 0& 1& 0& 1\cr
}\nonumber\end{eqnarray}}

Now, From eq. (\ref{indep}), the charge matrix for the
initial fields $X_i$ are given, via the charge matrix for the 
independent fields $x_b$ by the relation
\begin{equation}
\Delta=V\cdot K^t
\end{equation}
Hence, in this case, making use of the identity $U_{red}\cdot T^t_{red}
={\mbox{Id}}$, the matter content for the D-brane gauge theory on the  
resolved space is obtained as 
$\Delta_{red}=(VU)_{red}(T^t_{red}K^t_{red})$. 
From which in this case we obtain the matrix
{\small
\begin{eqnarray}
d_{red}=\pmatrix{
0&1&1&-1&-1&\cr
0&-1&-1&1&1\cr
}\label{compare1}\end{eqnarray}}
where we have included an extra row corresponding to the redundant 
$U(1)$.  Hence, the gauge group is seen to be $U(1)^2$, with four 
bifundamentals, in agreement with the result of \cite{pru}. Also note 
that one adjoint field has appeared in our expression. This signals an  
ambiguity that exists in the toric procedure with regards to handling
chargeless fields. Consider, for example, the 
superpotential interaction, which can be determined from the matrix
$K$ that encodes the F-flatness conditions. The number of columns in 
the matrix $K_{red}^t$ is the number of fields that appear in the 
interactions and the terms in the superpotential are read off as linear
relations between the columns of $K^t_{red}$. In this example, 
from the matrix $K$, we have the relation $X_2X_5=X_3X_4$, and
from the $d$ matrix, we can see that each of these have charge zero.
Hence, we try as an ansatz for the superpotential,
\begin{equation}
W=\left(X_1+\phi\right)\left(X_2X_5-X_3X_4\right)
\label{compare2}
\end{equation}
where $\phi$ is a field uncharged for both the $U(1)$'s.
This agrees with \cite{pru} after the identification
$x_{13}\rightarrow X_2;~Y_{34}\rightarrow X_5;~X_{24}
\rightarrow X_3;~X_{13}\rightarrow X_4;~Z_{41}\rightarrow X_1;~Z_{23}
\rightarrow -\phi$. As is clear, however, there is an ambiguity in
writing the superpotential due to the presence of fields that are 
chargeless under both the $U(1)$'s \cite{fhh}. 
Let us also mention that we can carry out the
inverse procedure just outlined for the case of the alternative
blowup to ${\bf C^3/Z_2\times C}$ shown in figure \ref{322figs}(d). 
This gives exactly the same matter content and superpotential as
the case just analysed, in the region of moduli space given by
$\zeta_1+\zeta_3=0$, $\zeta_2>>0$.\\

\noindent
$\bullet$
We now discuss the example of the blowup of the ${\bf C^3/Z_2\times
Z_2}$ singularity to the conifold. Proceeding in the same way as 
above, we now find that in the region of the FI parameters determined
by $\zeta_2=0;\zeta_{1,3}>>0$, the fields to be resolved are 
$[p_3,p_5,p_6,p_7,p_9]$, when the fields $[p_1,p_2,p_4,p_8]$ are chosen 
dynamical, i.e continue to have zero vev. Therefore, from eq. 
(\ref{qreduced}), we obtain the reduced charge matrix to be
\begin{equation}
Q_t^{red}=\left(-1~1~1~-1~\zeta_2\right) 
\end{equation}
Note that in this, case, there are no F-terms and we have 
$Q_{red}={\bf 0}_{1\times 4}$, moreover, the kernel of this
matrix is the $4\times 4$ identity matrix, i.e $T_{red}= 
{\mbox{Id}}_{4\times 4}$, which implies that (since the dual
matrix of $T_{red}$ is also the $4\times 4$ identity matrix)
$\Delta=Q_{red}$, with the charge matrix given in this case
(by adding the extra column corresponding to the overall $U(1)$)
as
{\small 
\begin{eqnarray}
d_{red}=\pmatrix{
-1&1&1&-1\cr
1&-1&-1&1\cr
}\label{d2}\end{eqnarray}}
In this case, the matrix $Q_{red}$ being zero, there are no F-terms
and the entire gauge theory information is contained in the D-term.
 
\section{D-branes on ${\bf C^3/Z_2\times Z_3}$ and its blowups}

Let us now consider D3-branes on the orbifold 
${\bf  C^3/Z_2 \times Z_3}$. In this case, the bosonic fields 
corresponding to the D3-brane coordinates
tangential to the ${\bf C^3}$ are $6\times 6$ matrices, and the
components that survive the projection by the quotienting group 
constitute the  $N=1$ matter multiplet, given by the set 
\begin{eqnarray}
\left(X_{12},X_{23},X_{34},X_{45},X_{56},X_{61}\right)\nonumber\\ 
\left(Y_{13},Y_{24},Y_{35},Y_{46},Y_{51},Y_{62}\right)\nonumber\\
\left(Z_{14},Z_{25},Z_{36},Z_{41},Z_{52},Z_{63}\right)
\label{323matter}
\end{eqnarray}
The F-flatness condition (\ref{superpot}) in this case imply $18$ 
constraints out of which $8$ are seen to be independent. Further,
there are $17$ homogeneous coordinates $p_{\alpha},~(\alpha=1\cdots 17)$, 
corresponding to the physical fields in terms of which the dual cone 
may be described. 

The total charge matrix of the homogeneous coordinates $p_{\alpha}$ is 
defined by introducing a set of $9$ ${\bf C^*}$ actions that remove 
the redundancy in expressing the original independent fields in terms 
of the homogeneous coordinates and a further set of $U(1)^5$ charges
expressing the D-term constraints in terms of the $p_{\alpha}$.   
(The $6$th $U(1)$ is redundant because of the relation
$\sum_i\zeta_i=0$). The total charge matrix can then be obtained
by concatenating these two sets of charge matrices as in our
earlier example, and is given (with the inclusion of a column 
specifying the $5$ independent Fayet-Illiopoulos parameters) by \cite{pru}
{\small
\begin{eqnarray}
Q_t=\pmatrix{
1&0&-1&0&0&0&0&0&0&0&-1&0&0&0&1&0&0&0\cr
0&1&0&0&0&0&-1&0&0&0&0&0&0&-1&0&0&1&0\cr
0&0&1&-1&0&0&0&0&0&-1&1&0&0&0&0&0&0&0\cr
0&0&0&1&-1&0&0&0&0&0&0&0&-1&1&0&0&0&0\cr
0&0&0&0&1&0&0&0&-1&0&-1&0&0&0&1&-1&1&0\cr
0&0&0&0&0&1&0&0&0&-1&0&0&-1&0&1&-1&1&0\cr
0&0&0&0&0&0&1&0&0&-1&-1&0&0&0&1&0&0&0\cr
0&0&0&0&0&0&0&1&-1&0&0&1&-1&0&-1&0&1&0\cr
0&0&0&0&0&0&0&0&0&0&0&1&0&-1&-1&1&0&0\cr
0&-1&0&0&1&0&0&0&0&0&0&0&0&0&0&0&0&\zeta_1\cr
-1&-1&0&0&0&1&1&0&0&0&0&0&0&0&0&0&0&\zeta_2\cr
1&0& -1&0&-1&0&0&0&1&0&0&0&0&0&0&0&0&\zeta_3\cr
0&1&0&1&-1&0&-1&0&0&0&0&0&0&0&0&0&0&\zeta_4\cr
-1&1&1&0&0&-1&0&0&0&0&0&0&0&0&0&0&0&\zeta_5\cr
}\nonumber\end{eqnarray} }
The toric data is the co-kernel of the transpose of this charge
matrix (when the FI parameters are set to zero) and is given by 
{\small
\begin{eqnarray}
{\tilde T}=\pmatrix{
1 & 0 & 1 & 1 & 0 & 0 & 1 & 1 & 0 & 1 & 1 & 0 & 0 & -1& 1 & 0 & 0 \cr
0 & 0 & 1 & -1& 0 & 1 & -1& 0 & 1 & 0 & -2& 0 & 0 & 1 & -1& 0 & 0 \cr
1 & 1 & 1 & 1 & 1 & 1 & 1 & 1 & 1 & 1 & 1 & 1 & 1 & 1 & 1 & 1 & 1 \cr
}\end{eqnarray}}

The invariant variables are defined by
\begin{eqnarray}
X_{12}X_{23}X_{34}X_{45}X_{56}X_{61}=x&=& p_1^4 p_2 p_3^6 p_4^2 p_5 p_6^3 
p_7^2 p_8^4 p_9^3 p_{10}^4 p_{12} p_{13} p_{15}^2 p_{16} p_{17}\nonumber\\
Y_{24}Y_{46}Y_{62}=y&=&p_1 p_2 p_4^2 p_5 p_7^2 p_8 p_{10} p_{11}^3 p_{12} 
p_{13} p_{15}^2 p_{16} p_{17} \nonumber\\ 
Z_{14}Z_{41}=z&=&p_2 p_5 p_6 p_9 p_{12} p_{13} p_{14}^2 p_{16} p_{17}\nonumber\\ 
X_{12}Y_{24}Z_{41}=w&=&p_1 p_2 p_3 p_4 p_5 p_6 p_7 p_8 p_9 p_{10} p_{11} 
p_{12} p_{13}\nonumber\\
&~&p_{14} p_{15} p_{16} p_{17}
\label{invariables}
\end{eqnarray}
From which it can be seen that 
\begin{equation}
xy^2z^3=w^6
\label{323eqn}
\end{equation}

\begin{figure}
\centering
\epsfxsize=4.5in
\hspace*{0in}\vspace*{.2in}
\epsffile{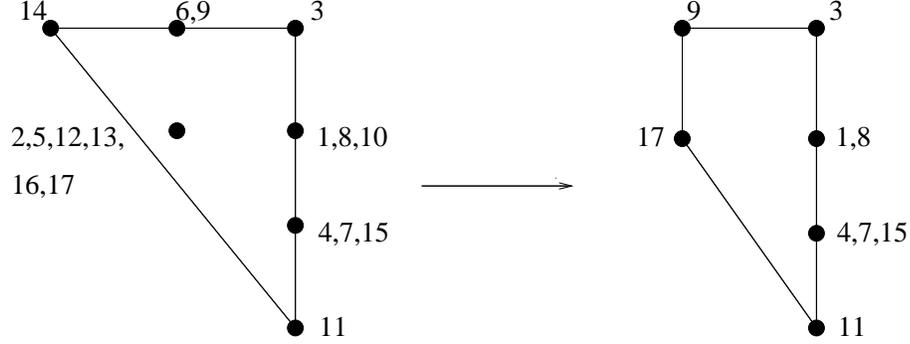}
\caption{\small Toric diagram showing a resolution of the
${\bf C^3/Z_2\times Z_3}$ orbifold along with the choice of fields
to be resolved.}
\label{3221}
\end{figure}
\noindent
Let us now discuss some examples of blowups of this space that will illustrate
the procedure outlined in section 2. We will use the notation $x_i=p_i^2$
in what follows.\\

\noindent
$\bullet$
Our first example is illustrated in 
figure \ref{3221}. From the diagram, we see that the field $p_{14}$
will definitely be resolved. Hence, after performing Gaussian row 
reduction on the charge matrix $Q_t$, which gives 
{\small
\begin{eqnarray}
\pmatrix{
1& 0& 0& 0& 0& 0& 0& 0& 0& 0& 1& 0& 0& 0& -2& 0& 0& -\zeta_1 - 
2\zeta_2 - \zeta_4 - 2\zeta_5\cr 
0& 1& 0& 0& 0& 0& 0& 0& 0& 0& 0& 0& 0& 0& 0& 0& -1& -\zeta_1 - 
\zeta_2 - \zeta_3\cr
0& 0& 1& 0& 0& 0& 0& 0& 0& 0& 2& 0& 0& 0& -3& 0& 0& -\zeta_1 - 
2\zeta_2 - \zeta_4 - 2\zeta_5\cr
0& 0& 0& 1& 0& 0& 0& 0& 0& 0& 0& 0& 0& 0& -1& 0& 0& -\zeta_2 - \zeta_5\cr
0& 0& 0& 0& 1& 0& 0& 0& 0& 0& 0& 0& 0& 0& 0& 0& -1& -\zeta_2 - \zeta_3\cr
0& 0& 0& 0& 0& 1& 0& 0& 0& 0& 1& 0& 0& 0& -1& 0& -1& -\zeta_1 
- \zeta_2 - \zeta_3 - \zeta_5\cr
0& 0& 0& 0& 0& 0& 1& 0& 0& 0& 0& 0& 0& 0& -1& 0& 0& -\zeta_1 - \zeta_2 
- \zeta_4 - \zeta_5\cr
0& 0& 0& 0& 0& 0& 0& 1& 0& 0& 1& 0& 0& 0& -2& 0& 0& -\zeta_2 - \zeta_5\cr
0& 0& 0& 0& 0& 0& 0& 0& 1& 0& 1& 0& 0& 0& -1& 0& -1& -\zeta_2\cr
0& 0& 0& 0& 0& 0& 0& 0& 0& 1& 1& 0& 0& 0& -2& 0& 0& -\zeta_1 - \zeta_2 
- \zeta_4 - \zeta_5\cr
0& 0& 0& 0& 0& 0& 0& 0& 0& 0& 0& 1& 0& 0& 0& 0& -1& \zeta_4 + \zeta_5\cr
0& 0& 0& 0& 0& 0& 0& 0& 0& 0& 0& 0& 1& 0& 0& 0& -1& \zeta_4\cr
0& 0& 0& 0& 0& 0& 0& 0& 0& 0& 0& 0& 0& 1& 1& 0& -2& -\zeta_3 + 
\zeta_4 + \zeta_5\cr
0& 0& 0& 0& 0& 0& 0& 0& 0& 0& 0& 0& 0& 0& 0& 1& -1& -\zeta_3\cr
}\label{grrqred}\end{eqnarray}} 

we use this matrix to solve for all the fields in terms of 
$x_{14}$, the solution set being given by
\begin{eqnarray}
\left[x_1\cdots x_{17}\right]&=&\left[x_1,x_2,(2x_1-2x_2+x_{14}-\zeta_1
-\zeta_3+\zeta_5),\right.\nonumber\\
&~&(2x_2-x_{14}+2\zeta_1 +\zeta_2+\zeta_3+\zeta_4),
(x_2+\zeta_1),\nonumber\\
&~&(x_1-x_2+x_{14}-\zeta_1-\zeta_3),(2x_2-x_{14}+\zeta_1
+\zeta_2+\zeta_3), \nonumber \\ 
&~&(x_1+\zeta_1+\zeta_2+\zeta_4+\zeta_5),
(x_1-x_2+x_{14}+\zeta_5),\nonumber\\
&~&(x_1+\zeta_2+\zeta_5),
(-x_1+4x_2-2x_{14}
+3\zeta_1+2\zeta_2+2\zeta_3+\zeta_4),\nonumber\\
&~&(x_2+\zeta_1+\zeta_2+\zeta_3
+\zeta_4+\zeta_5),
(x_2+\zeta_1+\zeta_2+\zeta_3+\zeta_4),\nonumber\\
&~&x_{14},(2x_2-x_{14}+2\zeta_1+2\zeta_2+\zeta_3+\zeta_4+\zeta_5),
\nonumber\\ &~&\left.(x_2+\zeta_1+\zeta_2),
(x_2+\zeta_1+\zeta_2+\zeta_3)\right]
\label{sol14}
\end{eqnarray}

Now, we choose the fields $[p_1,p_3,p_4,p_{11}]$ as the ones 
that continue to have zero vev, and in eq. (\ref{sol14}) set 
the values of these to zero. 
We thus obtain conditions on the FI parameters, $\zeta_1+\zeta_2+\zeta_4=0,~~
\zeta_2=2 \zeta_5$. In the new solution set obtained with these 
conditions, we choose $\zeta_5=0$ which makes $x_8=0$. This also imples 
that $\zeta_2=0$ (since $\zeta_2=2\zeta_5$). We also set $x_2=x_{14}=
-\zeta_3-\zeta_1$ to obtain $x_7,x_9,x_{15},x_{17}=0$. Our choice of 
$\zeta_2=\zeta_5=0$ implies that $x_{10}=0$. This gives the solution set
\begin{eqnarray}
\left[x_1\cdots x_{17}\right]&=&\left[0,(-\zeta_3-\zeta_1),0,0,-\zeta_3,
(-\zeta_1-\zeta_3),\right.\nonumber\\
&~&\left. 0,0,0,0,0,-\zeta_1,-\zeta_1,(-\zeta_3-\zeta_1),0,-\zeta_3,0\right]
\end{eqnarray}

Thus the fields to be resolved are 
$[p_2,p_5,p_6,p_{12},p_{13},p_{14},p_{16}]$. Hence, from (\ref{qreduced}), 
we obtain the reduced charge matrix,
{\small
\begin{eqnarray}
Q^{red}_t=\pmatrix{
0& 0& 1& 0& -1& 0& 0& -1& 1& 0& 0\cr 
0& 0& 0& 1& 0& 0& -1& -1& 1& 0& 0 \cr 
0& 1& -1& 0& 0& 0& -1& 1& 0& 0& 0\cr 
1& -1& 0& 0& 0& 0& 0& -1& 1& 0& 0\cr 
-1& 0& 1& 0& 0& -1& 1& -1& 0& 1& \zeta_2\cr 
0& 0& 1& -1& 0& 0& 0& 0& 0& 0& \zeta_1 + \zeta_4\cr 
-1& 1& -1& 1& 0& 1& -1& 1& 0& -1& \zeta_5 \cr 
}\nonumber\end{eqnarray}}

In the usual way, we calculate the kernel of the charge matrix $T_{red}$,
and the transpose of its dual matrix is
{\small\begin{eqnarray}
K_{red}^t=\pmatrix{
0& 1& 0& 0& 0& 0& 0& 0\cr 
0& 0& 1& 1& 0& 1& 0& 0\cr 
0& 0& 1& 1& 1& 0& 0& 0\cr 
0& 0& 0& 1& 0& 1& 1& 0\cr 
1& 0& 0& 0& 0& 0& 0& 0\cr 
0& 0& 1& 0& 0& 1& 0& 1\cr
}\nonumber\end{eqnarray}}

from the matrices $T_{red},K_{red}^t$ and  $(VU)_{red}$, 
we can read off the matter content and gauge group of the D-brane gauge
theory from the matrix
{\small\begin{eqnarray} d_{red} =\pmatrix{
1& -1& 0& 1& 0& 0& 0& -1\cr 
0& 0& 1& -1& 0& 0& -1& 1\cr 
-1& 1& 0& 0& -1& 1& 0& 0\cr
0& 0& -1& 0& 1& -1& 1& 0\cr
}\nonumber\end{eqnarray}}
This is an $U(1)^4$ gauge theory with $8$ matter
fields, in agreement with \cite{pru},\cite{ur}. To calculate 
the superpotential, we first note that the relations between the 
columns of the 
matrix $K^t$, which can be written as $X_3X_7=X_5X_6=X_4X_8$. All these
combinations are seen to be chargeless, as is the combination 
$X_1X_2$. Hence, our ansatz for the superpotential in this case is 
\begin{equation}
W=X_1X_2\left(X_3X_7-X_5X_6\right)+\phi_1\left(X_3X_7-X_4X_8\right)
+\phi_2\left(X_5X_6-X_4X_8\right)
\end{equation}
which agrees with that calculated in \cite{pru} after the 
identification with their notation, $Z_{25}\rightarrow X_1,
Z_{52}\rightarrow X_2;X_{23}\rightarrow X_3;Y_{62}\rightarrow X_7;
Y_{51}\rightarrow X_5;X_{45}\rightarrow X_6;\\Y\rightarrow X_4;
X_{61}\rightarrow X_8;Z_{36}\rightarrow -\phi_1; 
Z_{14}\rightarrow \phi_2$. 
 
In terms of the original variables of (\ref{323matter}), this
blowup corresponds to giving a vev to the fields $Z_{63},Z_{41}$.
The invariant variables are in this case defined by
\begin{eqnarray}
x'&=&p_1^2p_3^3p_4p_7p_8^2p_9p_{10}^2p_{15};~~~~~~x=x'^2z\nonumber\\
y&=&p_1p_4^2p_7^2p_8p_{10}p_{11}^3p_{15}^2p_{17}\nonumber\\
z&=&p_9p_{17}\nonumber\\
w'&=&p_1p_3p_4p_7p_8p_{10}p_{11}p_{15};~~~~~~w=w'z 
\end{eqnarray}
In terms of these variables, the blown up space is given by
$x'y=zw'^3$.\\

\begin{figure}
\centering
\epsfxsize=3.5in
\hspace*{0in}\vspace*{.2in}
\epsffile{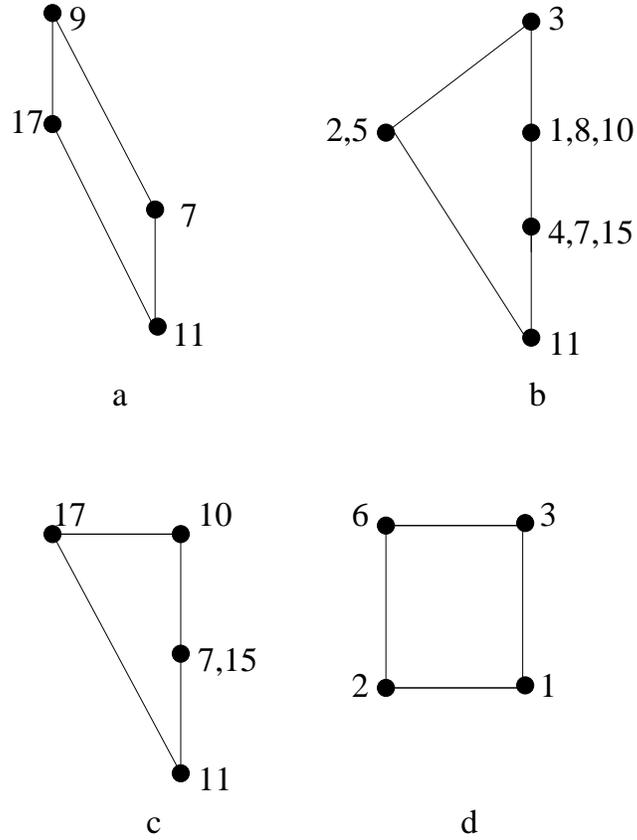}
\caption{\small Toric diagram showing some resolutions of the
${\bf C^3/Z_2\times Z_3}$ orbifold that we have considered. We have
marked the various fields that are chosen to remain dynamical on
resolving the parent singularity to these cases.}
\label{3222}
\end{figure}

\noindent
$\bullet$ We now consider the second example which is depicted in 
figure \ref{3222}(a). In this case, we start from the matrix
(\ref{grrqred}), and in the solution set of eq. (\ref{sol14}),
keeping in mind the toric diagram of the blowup, we select
the set of fields $[x_7,x_9]$ that remain dynamical, with
zero vev. This gives the solution $x_1=\zeta_1+\zeta_4$. 
From the resulting solution set, we now select $x_{17}=0$,
in order to obtain $x_2=x_{14}=-\zeta_1-\zeta_2-\zeta_3$ 
and after imposing this condition, we make the choice 
$\zeta_1+\zeta_4+\zeta_5=0$ in order for 
$p_9$ to have zero vev. The final solution is given by
\begin{eqnarray}
\left[x_1\cdots x_{17}\right]&=&\left[(\zeta_1+\zeta_4),
(-\zeta_1-\zeta_2-\zeta_3),(\zeta_1+\zeta_2+\zeta_4),
(\zeta_1+\zeta_4),\right .\nonumber\\
&~&(\zeta_2-\zeta_3),(\zeta_4-\zeta_3),0,(\zeta_1+\zeta_2+\zeta_4),0,
\zeta_2,0,-\zeta_1,\zeta_4, \nonumber\\
&~&\left .(-\zeta_1-\zeta_2-\zeta_3),\zeta_2,-\zeta_3,0\right]
\end{eqnarray}
Hence, the fields to be resolved are 
$[p_1,p_2,p_3,p_4,p_5,p_6,p_8,p_{10},p_{12},p_{13},p_{14},\\
p_{15},p_{16}]$ and the dynamical fields
are $[p_7,p_9,p_{11},p_{17}]$.
We can thus directly evaluate the reduced charge matrix from 
(\ref{qreduced}) which is given by  

\begin{equation}
Q_t^{red}=\left(-1~1~1~-1~\zeta_1 + \zeta_4 + \zeta_5\right)
\end{equation}

Defining the 
invariant variables from eq. (\ref{323matter}) in this case as
\begin{eqnarray}
x'&=&p_7p_9;~~~~~~x=x'^2z\nonumber\\
y'&=&p_{11}p_{17};~~~~~y=y'w'^2\nonumber\\
z&=&p_9p_{17}\nonumber\\
w'&=&p_7p_{11};~~~~~~w=w'z
\end{eqnarray}
this blownup space can be seen to be the conifold $x'y'=zw'$
by using the redefined variables in (\ref{323eqn}). The gauge
theory living on the D-brane is of course the same as in
eq. (\ref{d2}). In terms 
of the original variables in eq. (\ref{323matter}), this blowup 
can be shown to correspond to giving vevs to the fields $X_{45},
X_{23},Z_{63},Z_{41}$. Since there are no F-terms in this case, 
all the information about the gauge theory is contained in the D-term 
constraint.\\
 
\noindent
$\bullet$
Let us now consider the toric diagram given by figure \ref{3222}(b). 
From the toric diagram of fig. \ref{3222}(b), we choose 
the fields
$p_1,p_2,p_3,p_4,p_{11}$ to have zero vev. These conditions imply that
$\zeta_1+\zeta_4=0$ and also $\zeta_2+\zeta_5=0$. Substituting this in the 
solution set (\ref{sol14}), we obtain 
\begin{eqnarray}
\left[x_1\cdots x_{17}\right]&=&\left[0,0,0,0,\zeta_1,\zeta_2,0,0,(\zeta_1+
\zeta_3),0,0,\zeta_3,(\zeta_2+\zeta_3),\right. \nonumber\\
&~&\left. (\zeta_1+\zeta_3-\zeta_5),0,(\zeta_1+\zeta_2),
(\zeta_1+\zeta_2+\zeta_3)\right] 
\end{eqnarray}

Now, we have a choice. We put $\zeta_1=0$ and obtain $x_5=0$ which also implies
that $\zeta_4=0$. Hence, from the final solution set, we see that the 
fields that continue to have zero vev and are dynamical are, 
in this case, given by the set $[p_1,p_2,p_3,p_4,p_5,p_7,p_8,p_{10},
p_{11},p_{15}]$ and the fields that pick up non-zero vev and
hence are resolved are $[p_6,p_9,p_{12},p_{13},p_{14},p_{16},p_{17}]$. 
Hence, according to (\ref{qreduced}), we obtain the reduced
charge matrix as

{\small
\begin{eqnarray}
Q_t^{red}=\pmatrix{
0& 0& 0& 1& 0& 0& -1& 0& -1& 1& 0\cr
0& 0& 0& 0& 0& 1& 0& -1& -1& 1& 0\cr
0& 0& 1& -1& 0& 0& 0& -1& 1& 0& 0\cr
1& 0& -1& 0& 0& 0& 0& 0& -1& 1& 0\cr
-2& 0& 1& 0& 0& 1& 0& 0& 0& 0& \zeta_2 + \zeta_5\cr
0& 1& 0& 1& -1& -1& 0& 0& 0& 0& \zeta_4\cr
0& -1& 0& 0& 1& 0& 0& 0& 0& 0& \zeta_1\cr
}\nonumber\end{eqnarray}}
and the dual of the kernel of the charge matrix $Q_{red}$ is
{\small
\begin{eqnarray}K_{red}^t=\pmatrix{
1& 0& 0& 0& 0& 1& 1& 0\cr
1& 0& 0& 0& 0& 1& 0& 1\cr
1& 0& 0& 0& 1& 0& 1& 0\cr
0& 0& 0& 1& 0& 1& 1& 0\cr
0& 0& 1& 0& 0& 0& 0& 0\cr
0& 1& 0& 0& 0& 0& 0& 0\cr
}\nonumber\end{eqnarray}}
As before, in order to calculate the matter content, we use the matrix 
$T_{red}^t\cdot K_{red}^t$, from which the matrix $\Delta$, which specifies 
the matter content and gauge group of the D-brane gauge theory is 
calculated to be
{\small\begin{eqnarray}
d_{red}=\pmatrix{
1& 0& 0& -1& 0& 0& 1& -1\cr
-1& 1& -1& 1& -1& 1& 0& 0\cr
0& -1& 1& 0& 0& 0& 0& 0\cr
0& 0& 0& 0& 1& -1& -1& 1\cr
}\nonumber\end{eqnarray}}
Hence, the gauge group is $U(1)^4$ and there are $8$ matter fields. 
Using the relations $X_1X_4=X_5X_6=X_7X_8$ with total charge zero and 
the chargeless combination $X_2X_3$, we write the superpotential in 
this case as 
\begin{equation}
W=X_2X_3\left(X_1X_4-X_5X_6\right)+\phi_1\left(X_1X_4-X_7X_8\right)
+\phi_2\left(X_5X_6-X_7X_8\right)
\end{equation}
where the fields $\phi_1,\phi_2$ are possible adjoints.
The invariant variables (\ref{invariables}) are 
defined by
\begin{eqnarray}
x'&=&p_1^2p_3^3p_4p_7p_8^2p_{10}^2p_{15}\nonumber\\
y'&=&p_1p_4^2p_7^2p_8p_{10}p_{11}^3p_{15}^2\nonumber\\
z&=&p_2p_5\nonumber\\
w'&=&p_1p_3p_4p_7p_8p_{10}p_{11}p_{15}
\end{eqnarray}
with the relations $x=x'^2z;~y=y'z;~w=w'z$. The remaining singularity
is seen from eq. (\ref{323eqn}) to be the equation 
$x'y'=w'^3$ in ${\bf C^3}$, which is recognised
as corresponding to blowing up the ${\bf C^3/Z_2\times Z_3}$ orbifold
to the singularity ${\bf C^2/Z_3 \times C}$. Let us also mention 
that in terms of the original variables in eq. (\ref{323matter}),
this blowup corresponds to giving vevs to the fields $Y_{13},
Z_{14},Z_{52},Z_{63}$.\\

\noindent
$\bullet$
We now come to the example shown in figure \ref{3222}(c). 
In this case, from the Gaussian row-reduced matrix 
of eq. (\ref{sol14}), we choose
the set of fields $[p_{10},p_7,p_{11}]$ to have zero vev. This implies
that $\zeta_1+\zeta_2+\zeta_4+\zeta_5=0$ and also that $x_1=\zeta_1+\zeta_4$.
Substituting this in the solution set implies that in addition $x_{15}=0$, 
and further setting $x_2=-\zeta_3+\zeta_4+\zeta_5$, so that $x_{17}$ is zero, 
we obtain the full solution set in this case as
\begin{eqnarray}
\left[x_1\cdots x_{17}\right]&=&\left[(-\zeta_2-\zeta_5),(-\zeta_3+\zeta_4+\zeta_5),
(-\zeta_2-\zeta_5),(-\zeta_2-\zeta_5),\right. 
\nonumber\\&~&(-\zeta_2-\zeta_3),(\zeta_4-\zeta_5),0,(-\zeta_2-\zeta_5),
-\zeta_2,0,0, (\zeta_4+\zeta_5),\zeta_4, \nonumber\\
&~&\left.(-\zeta_3+\zeta_4+\zeta_5),0,-\zeta_3,0\right]
\end{eqnarray}  
 
Hence, the fields that are dynamical and have zero vev is given by the
set $[p_7,p_{10},p_{11},p_{15},p_{17}]$ and the set to be resolved is
given by $[p_1,p_2,p_3,p_4,p_5,p_6,\\p_8,p_9,p_{12},p_{13},p_{14},p_{16}]$.
Using this, the reduced charge matrix is calculated from (\ref{qreduced})
to be  
{\small\begin{eqnarray}
Q_t^{red}=\pmatrix{
1&-1&-1&1&0&0\cr
0&-1&-1&2&0&\zeta_1+\zeta_2+\zeta_4+\zeta_5\cr
}\nonumber\end{eqnarray}}

From which the F and the D-terms as 

\begin{equation}
Q_{red}=\left(1~-1~-1~1~0\right)~~~~~~(VU)_{red}=\left(0~-1~-1~2~0\right)
\end{equation}
The dual  of the kernel of $Q_{red}$ is given by 
{\small\begin{eqnarray}
K_{red}^t=\pmatrix{
1& 0& 0& 0& 0\cr 
0& 1& 1& 0& 0\cr
0& 1& 0& 1& 0\cr
0& 0& 1& 0& 1\cr
}\nonumber\end{eqnarray}}

and the matter content and gauge group of the D-brane gauge theory 
is obtained as

{\small\begin{eqnarray}
d_{red}=\pmatrix{
0&1&1&-1&-1\cr
0&-1&-1&1&1\cr
}\nonumber\end{eqnarray}}
The superpotential is written down, after noting that the field
$X_1$ is chargeless, and the relations $X_2X_5=X_3X_4$ from the
matrix $K^t$. It is given by
\begin{equation}
W=\left(X_1+\phi\right)\left(X_2X_5-X_3X_4\right)
\end{equation}
where $\phi$ is a possible adjoint.
The matter content and the superpotential are exactly 
the same as in (\ref{compare1}) and (\ref{compare2}). This is of
course as expected because both describe the world volume theory
of a D3-brane on the ${\bf C^2/Z_2}$ singularity. 
This blowup corresponds to giving vevs to the fields
$X_{12},X_{23},X_{45}$. The invariant variables (\ref{invariables})
are in this case
\begin{eqnarray}
x'&=&p_7p_{10}^2p_{15}\nonumber\\
y'&=&p_7p_{11}^2p_{15}\nonumber\\
z&=&p_{17}\nonumber\\
w'&=&p_7p_{10}p_{11}p_{15}
\end{eqnarray}
where $x=x'^2z;y=y'w'z;w=w'z$. The blown up space is, from eq. 
(\ref{323eqn}), given by the equation $x'y'=w'^2$ in ${\bf C^3}$, 
which describes the blowup 
to ${\bf C^2/Z_3 \times C}$. Let us pause to comment here that
the toric diagram of figure \ref{3222}(c) can be embedded both
into that of ${\bf C^3/Z_2\times Z_3}$ and ${\bf C^3/Z_2\times Z_2}$;
in fact, this diagram is identical to that in figure \ref{322figs}(d).
The fact that either embedding gives correct results for the 
D-brane gauge theory data shows that the inverse toric procedure is
indeed consistent.\\ 

\noindent
$\bullet$ 
Finally, let us discuss the toric diagram of figure \ref{3222}(d).
which is the conifold obtained by partially resolving the 
${\bf C^2/Z_2\times Z_3}$ singularity.
Following the same procedure as outlined in the previous examples,
we start from the solution set of (\ref{sol14}) and
select the fields $[p_1,p_3,p_6]$ which
we set to zero. From this, we obtain the solution $x_2=\zeta_5$
and $x_{14}=(\zeta_1+\zeta_2+\zeta_5)$. Substituting this in
the solution set and further setting $\zeta_5=0$ so that $x_2=0$,
we finally obtain the full solution set to be
\begin{eqnarray}
\left[x_1\cdots x_{17}\right]&=&\left[0,0,0,(\zeta_1+\zeta_2+\zeta_4),
\zeta_1,0,\zeta_2,(\zeta_1+\zeta_2+\zeta_3+\zeta_4),
(\zeta_1+\zeta_3),\right. \nonumber\\
&~&\zeta_2,(\zeta_1+2\zeta_2+\zeta_4)(\zeta_1+\zeta_2+\zeta_3+\zeta_4),
(\zeta_1+\zeta_2+\zeta_3 +\zeta_4),\nonumber\\
&~&\left.(\zeta_1+\zeta_3),(\zeta_1+2\zeta_2+\zeta_4),(\zeta_1+\zeta_2),
(\zeta_1+\zeta_2+\zeta_3)\right]
\end{eqnarray}
Hence, the fields that are dynamical are given by the set
$[p_1,p_2,p_3,p_6]$ while the rest acquire non-zero vev and hence
are resolved, and in this case the reduced charge matrix is obtained 
from (\ref{qreduced}) to be
\begin{equation}
Q=\left(-1~~1~~1~~-1~~\zeta_5\right)
\end{equation}

This blowup corresponds to the conifold. To see this, we define the
invariant variables (\ref{invariables}) in this case by
\begin{equation}
x'=p_1p_3=\frac{w}{z};~~~~y=p_1p_2;~~~~z=p_2p_6;~~~~w'=p_3p_6
\end{equation}
In this case, from (\ref{323eqn}), we can see that the remaining
singularity is given by the conifold $x'y=zw'$. Let us also mention that
in terms of the original variables in eq. (\ref{323matter}), this 
blowup corresponds to giving vevs to $Y_{13},Y_{24},Z_{36}$. 

\section{D-branes on ${\bf C^4/Z_2\times Z_2\times Z_2}$ and its
blowups}

We now consider D-branes on the orbifold ${\bf C^4/Z_2\times Z_2\times Z_2}$
and some of its blowups \cite{ak}. We will consider a D1-brane at this
singularity, and the world volume theory is an $N=(0,2)$ SYM theory in
two dimensions. The method of construction of the moduli space of the 
D-1 brane is similar to the method outlined in sections 2 and 3.  
The unbroken gauge symmetry in this case is (apart from a redundant
$U(1)$), $U(1)^7$. 
The matter multiplet, after imposing the appropriate projections 
consist of $32$ surviving fields, 
\begin{eqnarray}
\left(X_{18},X_{27},X_{36},X_{45},X_{54},X_{63},X_{72},X_{81}\right),
\left(Y_{15},Y_{26},Y_{37},Y_{48},Y_{51},Y_{62},Y_{73},Y_{84}\right)
\nonumber\\
\left(Z_{13},Z_{24},Z_{31},Z_{42},Z_{57},Z_{68},Z_{75},Z_{86}\right),
\left(W_{12},W_{21},W_{34},W_{43},W_{56},W_{65},W_{78},W_{87}\right)
\end{eqnarray}
The F-term constraints of eq. (\ref{superpot}), in this case imply 
that out of the $32$ fields, 
$11$ are independent, and the $32$ initial fields
can be solved in terms of these $11$, and the solutions can be
denoted as vectors in the lattice ${\bf Z^{11}}$. The dual cone
in this case has $34$ homogeneous coordinates (matter fields) $p_{\alpha}$, 
and the redundancy in the definition of the initial independent 
coordinates in terms of these can be eliminated by introducing a set of 
$23$ ${\bf C^*}$ actions. The charge matrix $Q$ for the ${\bf C^*}$ actions
is given in \cite{ak}. For the matrix denoting the original $U(1)$ charges
expressed in terms of the homogeneous coordinates $p_{\alpha}$, we use
the Moore-Penrose inverse of the matrix defining the dual cone, and
concatenate these two matrices to obtain the total charge matrix given
in the appendix. 
The toric diagram is given by the columns of the matrix 
\[
{\small
\begin{array}{l}
\left(
\matrix{
p_1&p_2&p_3&p_4&p_5&p_6&p_7&p_8&p_9&p_{10}&p_{11}&p_{12}&p_{13}&
p_{14}& p_{15}&p_{16}&p_{17}\cr
1& 0& 1& 0& 1& 2& 1& 0& 0& 1& 0& 1& 1& 1& 1& 1& 1\cr 
1& 1& 1& 1& 0& 0& 0& 0& 0& 0& 2& 1& 0& 1& 1& 1& 1\cr 
1& 0& 0& 1& 1& 0& 0& 1& 2& 1& 0& 0& 0& 1& 1& 1& 1\cr 
0& 0& 0& -1& 0& 1& 1& 0& -1& 0& -1& 0& 1& 0& 0& 0& 0
\cr}
\right. \cdots \cdots
\\ \\ \\
\left.
\cdots\matrix{
p_{18}&p_{19}&p_{20}&p_{21}&p_{22}&p_{23}&p_{24}&p_{25}&p_{26}&
p_{27}&p_{28}&p_{29}&p_{30}&p_{31}&p_{32}&p_{33}&p_{34}\cr
1& 0& 0& 0& 1& 1& 1& 1& 1& 1& 0& 1& 1& 1& 1& 1& 1\cr 
1& 1& 0& 0& 1& 1& 1& 1& 1& 1& 1& 1& 1& 1& 1& 1& 1\cr 
1& 1& 0& 1& 1& 1& 1& 1& 1& 1& 0& 1& 1& 1& 1& 1& 1\cr 
0& -1& 1& 0& 0& 0& 0& 0& 0& 0& 0& 0& 0& 0& 0& 0& 0
\cr}
\right)
\end{array}
}
\]
\begin{figure}
\centering
\epsfxsize=5.5in
\hspace*{0in}\vspace*{.2in}
\epsffile{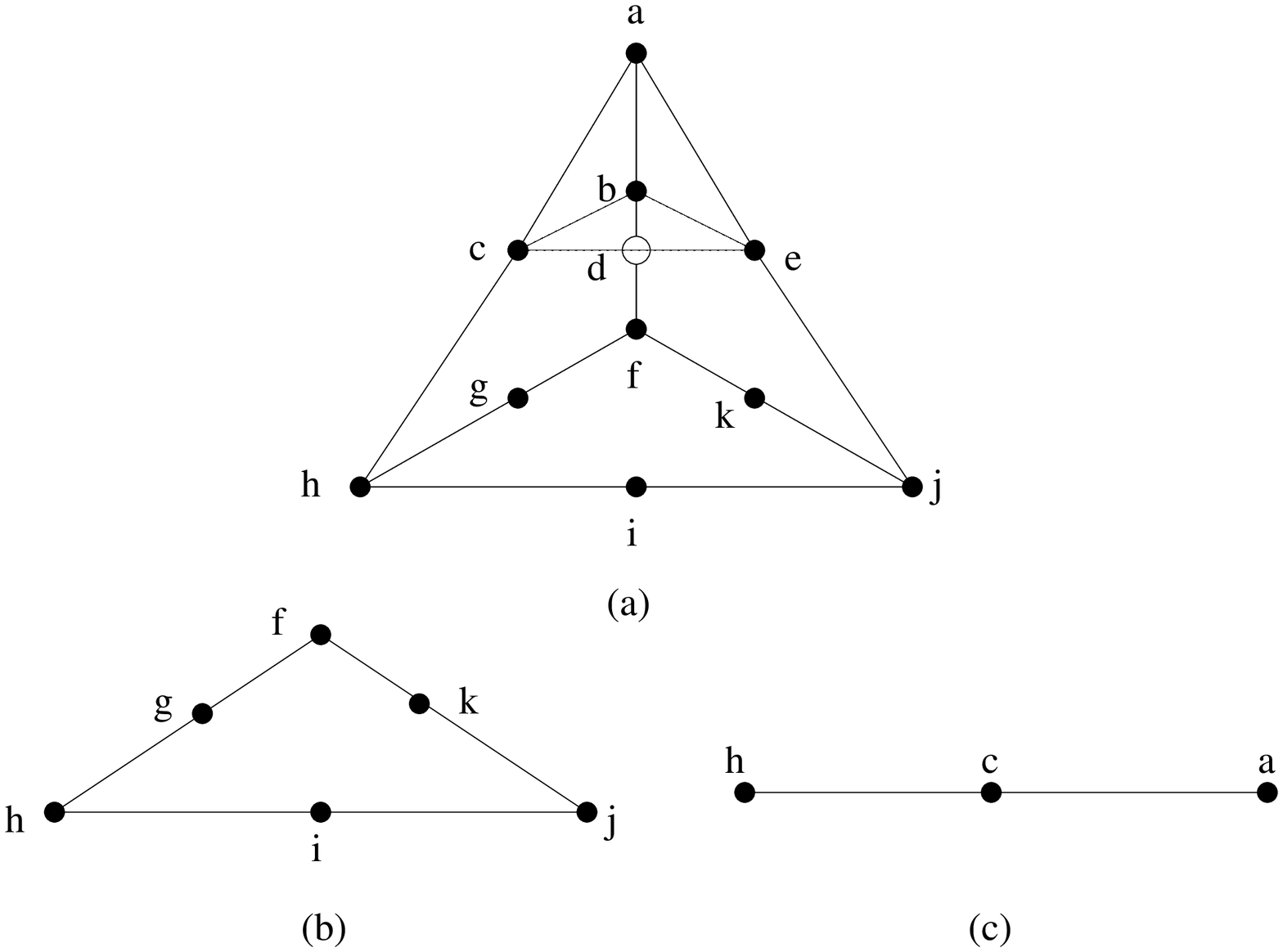}
\caption{\small Toric data for the orbifold ${\bf C^4/Z_2\times Z_2\times Z_2}
$ and two of its blowups to the ${\bf Z_2\times Z_2}$ singularity and the
${\bf Z_2}$ singularity.}
\label{4222}
\end{figure}
The distinct columns in the toric data are shown in fig. \ref{4222}.
The labelling of the vectors and the corresponding
$p_{\alpha}$ are as follows
\begin{eqnarray}
a&~&~(2~0~0~1)~~~~~~~[p_6]\nonumber\\
b&~&~(1~0~0~1)~~~~~~~[p_7,p_{13}]\nonumber\\
c&~&~(1~0~1~0)~~~~~~~[p_5,p_{10}]\nonumber\\
d&~&~(1~1~1~0)~~~~~~~[p_1,p_{14},p_{15},p_{16},p_{17},p_{18},p_{22},p_{23},
\nonumber\\
&~&~~~~~~~~~~~~~~~~~~~~~~p_{24},p_{25},p_{26},p_{27},p_{29},p_{30},p_{31},p_{32},
p_{33},p_{34}]\nonumber\\
e&~&~(1~1~0~0)~~~~~~~[p_3,p_{12}]\nonumber\\
f&~&~(0~0~0~1)~~~~~~~[p_{20}]\nonumber\\
g&~&~(0~0~1~0)~~~~~~~[p_8,p_{21}]\nonumber\\
h&~&~(0~0~2~-1)~~~[p_9]\nonumber\\
i&~&~(0~1~1~-1)~~~[p_4,p_{19}]\nonumber\\
j&~&~(0~2~0~-1)~~~[p_{11}]\nonumber\\
k&~&~(0~1~0~0)~~~~~~~[p_{11}]
\end{eqnarray}
The invariant variables, in this case denoted by
$x=(X_{18}X_{81})$; $y=(Y_{15}Y_{51})$;\\
$z=(Z_{13}Z_{31})$; $w=(W_{12}W_{21})$;
$v=(X_{18}Y_{84}Z_{42}W_{21})$, are 
\begin{eqnarray}
x&=&p_1p_2p_7p_8p_{13}p_{14}p_{15}p_{16}p_{17}p_{18}
p_{20}^2p_{21}p_{22}p_{23}p_{24}\nonumber\\
&~&p_{25}p_{26}p_{27}p_{28}p_{29}p_{30}
p_{31}p_{32}p_{33}p_{34}\nonumber\\
y&=&p_1p_2p_3p_4p_{11}^2p_{12}p_{14}p_{15}p_{16}p_{17}
p_{18}p_{19}p_{22}p_{23}p_{24}\nonumber\\
&~&p_{25}p_{26}p_{27}p_{28}p_{29}p_{30}
p_{31}p_{32}p_{33}p_{34}\nonumber\\
z&=&p_1p_4p_5p_8p_9^2p_{10}p_{14}p_{15}p_{16}p_{17}
p_{18}p_{19}p_{21}p_{22}p_{23}p_{24}\nonumber\\
&~&p_{25}p_{26}p_{27}p_{29}p_{30}
p_{31}p_{32}p_{33}p_{34}\nonumber\\
w&=&p_1p_3p_5p_6^2p_7p_{10}p_{12}p_{13}p_{14}p_{15}
p_{16}p_{17}p_{18}p_{22}p_{23}p_{24}\nonumber\\
&~&p_{25}p_{26}p_{27}p_{29}p_{30}
p_{31}p_{32}p_{33}p_{34}\nonumber\\
v&=&p_1^2p_2p_3p_4p_5p_6p_7p_8
p_9p_{10}p_{11}p_{12}p_{13}p_{14}^2p_{15}^2p_{16}^2
p_{17}^2\nonumber\\
&~&p_{18}^2p_{19}p_{20}p_{21}p_{22}^2
p_{23}^2p_{24}^2 p_{25}^2p_{26}^2p_{27}^2p_{28}p_{29}^2
p_{30}^2p_{31}^2p_{32}^2 p_{33}^2p_{34}^2
\label{inv422}
\end{eqnarray}

In terms of these variables, the space is defined by the
surface $xyzw=v^2$ in ${\bf C^5}$. 
Two of its resolutions (corresponding to cases 
(Vb and VIb of \cite{ak}) are shown in figure \ref{4222}.\\ 
This singularity has to be analysed in the lines of \cite{bglp}
in order to determine whether its partial resolutions are 
realised in the moduli space of the D-brane world volume gauge
theory. We leave this issue for future work, and for the moment
analyse the blowups of this singularity into lower dimensional
orbifold singularities, using the inverse toric procedure.\\
 
\noindent
$\bullet$
Let us consider the blowup illustrated in fig. \ref{4222}(b).
Performing Gaussian row reduction on the total charge matrix, we find
that for the range of the FI parameters $\zeta_1+\zeta_2=0;~\zeta_3
+\zeta_4=0;~\zeta_5+\zeta_6=0$, an appropriate initial choice 
determines the fields that retain zero vev 
as the set $[p_2,p_4,p_8,p_9,p_{11},p_{19},
p_{20},p_{21},p_{28}]$, while all others have non-zero vev and are hence
resolved. The reduced charge matrix is  
{\small\begin{eqnarray}
Q_t^{red}=\pmatrix{
0& 1& 0& -1& -1& 1& 0& 0& 0& 0\cr 
-1& 0& 0& 0& 1& 0& 1& 0& -1& 0\cr 
-1& 0& 1& -1& 1& 0& 0& 1& -1& 0\cr
5& 5& -5& 0& -4& -1& 0& 1& -1&\zeta_5 + \zeta_6\cr 
-5& 5& 5& -4& 0& -1& 0& -1& 1&\zeta_3 + \zeta_4\cr 
-1& -5& -5& 4& 0& 1& 0& 1& 5&\zeta_1 + \zeta_2\cr 
}\nonumber\end{eqnarray}}
From the kernel of the charge matrix $Q_{red}$, the transpose of its
dual is obtained as
{\small\begin{eqnarray}
K^t=\pmatrix{
1& 1& 0& 1& 0& 1& 0& 0& 0& 0& 0& 0\cr 
1& 0& 1& 0& 0& 0& 1& 0& 1& 0& 0& 0\cr 
1& 0& 0& 1& 0& 0& 0& 0& 1& 0& 1& 0\cr 
0& 1& 1& 0& 1& 0& 0& 0& 0& 1& 0& 0\cr 
0& 1& 0& 0& 0& 1& 0& 0& 0& 1& 0& 1\cr 
0& 0& 1& 0& 1& 0& 1& 1& 0& 0& 0& 0\cr
}\nonumber\end{eqnarray}}

From these matrices, we can extract the gauge theory data which
is now given by the matrix 
{\small\begin{eqnarray}
d_{red}=\pmatrix{
0& -1& 0& -1& -1& 0& 1& 0& 1& 0& 0& 1\cr
0& 0& -1& 1& 0& 1& 0& 1& -1& -1& 0& 0 \cr
1& 1& 1& 0& 0& 0& 0& -1& 0& 0& -1& -1\cr
-1&0& 0& 0& 1&-1&-1& 0& 0& 1& 1& 0& \cr
}\nonumber\end{eqnarray}}
The gauge group is as expected, $U(1)^4$ with $12$ matter fields, 
and the superpotential is given by
\begin{eqnarray}
W&=&X_1X_5X_{12}-X_3X_4X_{12}+X_2X_8X_9-X_2X_7X_{11}+X_3X_6X_{11}
\nonumber \\
&-&X_5X_6X_9+X_{10}X_4X_7-X_1X_8X_{10}
\end{eqnarray}
From the definition of the invariant variables in eq. (\ref{inv422}), it can
be seen that this space is ${\bf C^3/Z_2\times Z_2}$ defined by the equation
$xyz=v^2$ in ${\bf C^4}$. In terms of the original fields, it corresponds
to giving vevs to the fields $W_{12}$ and $W_{21}$.\\ 

\noindent
$\bullet$
Finally, we come to the example of fig. \ref{4222}(c). In this case,
in the region of moduli space given by $\zeta_2+\zeta_3+\zeta_6
+\zeta_7=0$, we find that the dynamical fields are given by the
set $[p_5,p_6,p_9,p_{10}]$, while the rest acquire non-zero vev and
are resolved. The reduced charge matrix is 
{\small\begin{eqnarray}
Q_t^{red}=\pmatrix{
1& -1& -1& 1& 0\cr
5& -2& -2& -1& \zeta_2+\zeta_3+\zeta_6+\zeta_7\cr
}\nonumber\end{eqnarray}}
The matrix $K^t$ is in this case given by
{\small\begin{eqnarray}
K^t=\pmatrix{
1&1&0&0\cr
1&0&1&0\cr
0&1&0&1\cr
}\nonumber\end{eqnarray}}
Which specifies the matter content as a $U(1)^2$ gauge theory with
the charge matrix given, as expected, by
{\small\begin{eqnarray}
d_{red}=\pmatrix{
-1& -1& 1& 1\cr
1& 1& -1&  -1\cr
}\nonumber\end{eqnarray}}
With a superpotential $W=\phi\left(X_1X_4-X_2X_3\right)$ with $\phi$
being a possible chargeless field. In terms of the original fields in
the theory, this resolution corresponds to giving vevs to $X_{18},X_{81},
Y_{15},Y_{51}$, and the resulting space is the ${\bf C^2/Z_2}$ orbifold
$zw=v^2$ that we have discussed earlier.

\section{Discussions}

In this paper, we have critically examined the inverse procedure of 
obtaining the world volume gauge theory data of D-branes 
probing certain toric singularities and their blowups from the geometric 
data of the resolution. We have shown
that the algorithm of \cite{fhh} gives consistent results, for a given
singularity that can be reached by partial resolutions of different
parent theories. We have explicitly checked several partial resolutions of 
orbifolds of the form ${\bf C^3}/\Gamma$ and ${\bf C^4}/\Gamma$ and 
found the results to be in agreement with field theory calculations. 
However, as pointed out 
in \cite{fhh}, in the presence of chargeless fields in the matter 
multiplet, the procedure cannot be used to specify the superpotential 
interaction uniquely, because of the inherent problem in the handling of 
such fields by toric methods.  

We have treated the simplest examples of the resolution of the\\ 
${\bf C^4/Z_2\times Z_2\times Z_2}$ singularity in this paper.
It would be interesting to investigate the various partial resolutions of 
the moduli space of D-branes on this singularity 
the lines of \cite{bglp} and determine the gauge theory data thereof.

Finally, as noted in \cite{fhh}, the ambiguities that exist in the 
inverse algorithm seem to imply that in some cases, different gauge theories,
in the infra red limit flow to theories with identical moduli space.  
We have not dealt with this aspect in the present paper, and it would be
an interesting direction for future work.

\vskip0.4cm
\begin{center}
{\bf Acknowledgements}
\end{center}
I would like to thank I. Biswas, S. Bhattacharjee, P. Chatterjee
and D. Suryaramana for helpful discussions.

\pagebreak
{\large {\bf Appendix}}
\vskip0.2cm
{\bf Charge matrix for the partial resolutions of ${\bf C^3/Z_2\times Z_2\times
Z_2}$:}
\[
{\tiny
\begin{array}{l}
Q_t =
\left(
\matrix{ p_{1}& p_{2}& p_{3}& p_{4}& p_{5}& p_{6}& p_{7}& p_{8}& p_{9}& 
p_{10}& p_{11}& p_{12}& p_{13}& p_{14}& p_{15}& p_{16}& p_{17}&\cr
0& 0& 0& 0& 1& -1& 0& 0& -1& 1& 0& 0& 0& 0& 0& 0& 0\cr 
0& 0& 1& 0& 0& -1& 0& 0& 0& 0& -1& 1& 0& 0& 0& 0& 0\cr 
-1& 0& 0& 0& 1& 0& 0& 1& -1& 0& 0& 0& -1& 1& 0& 0& 0\cr 
1& -1& 0& -1& -1& 0& -1& 0& 0& 0& 0& 0& 0& 0& 1& 0& 0\cr 
1& -1& -1& 0& -1& 0& 0& -1& 0& 0& 0& 0& 0& 0& 0& 1& 0\cr 
1& -1& 0& -1& 0& -1& 0& -1& 0& 0& 0& 0& 0& 0& 0& 0& 1\cr 
-1& 0& 1& 1& 1& -1& 0& 0& -1& 0& -1& 0& 0& 0& 0& 0& 0\cr 
0& 0& 0& 1& 0& 0& 0& 0& -1& 0& -1& 0& 0& 0& 0& 0& 0\cr 
0& 0& 0& 0& 0& 1& -1& 0& 0& 0& 0& 0& -1& 0& 0& 0& 0\cr 
0& 0& 0& 0& 0& 1& -1& 1& -1& 0& 0& 0& -1& 0& 0& 0& 0\cr 
-2& 1& 1& 1& 1& 0& 0& 1& -1& 0& -1& 0& -1& 0& 0& 0& 0\cr 
1& 0& -1& -1& 0& 0& -1& -1& 0& 0& 0& 0& 0& 0& 0& 0& 0\cr 
1& -1& -1& 0& 0& 0& -1& 0& -1& 0& 0& 0& 0& 0& 0& 0& 0\cr 
-1& 1& 1& 0& 0& 0& 0& 0& 0& 0& -1& 0& -1& 0& 0& 0& 0\cr 
-1& 0& 1& 1& 0& 0& 0& 1& -1& 0& -1& 0& -1& 0& 0& 0& 0\cr 
2& -1& -1& -1& -1& 0& -1& -1& 0& 0& 0& 0& 0& 0& 0& 0& 0\cr 
0& 1& 0& 0& 0& 1& -1& 0& 0& 0& -1& 0& -1& 0& 0& 0& 0\cr 
1& 0& -1& -1& -1& 1& -1& 0& 0& 0& 0& 0& -1& 0& 0& 0& 0\cr 
-1& 1& 0& 1& 0& 1& -1& 1& -1& 0& -1& 0& -1& 0& 0& 0& 0\cr 
-1& 1& 1& 0& 1& 0& -1& 1& -1& 0& -1& 0& -1& 0& 0& 0& 0\cr 
-1& 1& 0& 1& 1& 0& 0& 0& -1& 0& -1& 0& -1& 0& 0& 0& 0\cr 
1& 0& 0& 0& -1& 0& -1& -1& 0& 0& -1& 0& 0& 0& 0& 0& 0\cr 
1& 0& 0& 0& 0& 0& -1& 0& -1& 0& -1& 0& -1& 0& 0& 0& 0\cr 
-5& -2& -2& -2& -2& 0& -2& -2& 0& 2& 0& 2& 2& 1& -3& -3& -1\cr 
-1& -2& 2& -2& 2& 0& 2& -2& 0& -2& 0& -2& -2& -3& 1& 1& -5\cr 
-1& -2& -2& 2& 2& 0& -2& 2& 0& -2& 0& 2& 2& -3& 1& 1& 3\cr 
3& -2& 2& 2& -2& 0& 2& 2& 0& 2& 0& -2& -2& 1& -3& -3& -1\cr 
-1& 2& 2& 2& -2& 0& -2& -2& 0& 2& 0& -2& 2& 5& 1& 1& 3\cr 
3& 2& -2& 2& 2& 0& 2& -2& 0& -2& 0& 2& -2& 1& 5& -3& -1\cr 
3& 2& 2& -2& 2& 0& -2& 2& 0& -2& 0& -2& 2& 1& -3& 5& -1\cr}
\right. \cdots \cdots
\\ \\ \\
\left.
\cdots\matrix{
 p_{18}& p_{19}& p_{20} & p_{21}& p_{22}& p_{23}& p_{24}&p_{25}& 
p_{26}& p_{27}& p_{28}& p_{29}& p_{30}& p_{31}& p_{32}& p_{33}& p_{34}
& p_{35}& \cr
0& 0& 0& 0& 0& 0& 0& 0& 0& 0& 0& 0& 0& 0& 0& 0& 0& 0\cr 
0& 0& 0& 0& 0& 0& 0& 0& 0& 0& 0& 0& 0& 0& 0& 0& 0& 0\cr 
0& 0& 0& 0& 0& 0& 0& 0& 0& 0& 0& 0& 0& 0& 0& 0& 0& 0\cr 
0& 0& 0& 0& 0& 0& 0& 0& 0& 0& 0& 0& 0& 0& 0& 0& 0& 0\cr 
0& 0& 0& 0& 0& 0& 0& 0& 0& 0& 0& 0& 0& 0& 0& 0& 0& 0\cr 
0& 0& 0& 0& 0& 0& 0& 0& 0& 0& 0& 0& 0& 0& 0& 0& 0& 0\cr 
1& 0& 0& 0& 0& 0& 0& 0& 0& 0& 0& 0& 0& 0& 0& 0& 0& 0\cr 
0& 1& 0& 0& 0& 0& 0& 0& 0& 0& 0& 0& 0& 0& 0& 0& 0& 0\cr 
0& 0& 1& 0& 0& 0& 0& 0& 0& 0& 0& 0& 0& 0& 0& 0& 0& 0\cr 
0& 0& 0& 1& 0& 0& 0& 0& 0& 0& 0& 0& 0& 0& 0& 0& 0& 0\cr 
0& 0& 0& 0& 1& 0& 0& 0& 0& 0& 0& 0& 0& 0& 0& 0& 0& 0\cr 
0& 0& 0& 0& 0& 1& 0& 0& 0& 0& 0& 0& 0& 0& 0& 0& 0& 0\cr 
0& 0& 0& 0& 0& 0& 1& 0& 0& 0& 0& 0& 0& 0& 0& 0& 0& 0\cr 
0& 0& 0& 0& 0& 0& 0& 1& 0& 0& 0& 0& 0& 0& 0& 0& 0& 0\cr 
0& 0& 0& 0& 0& 0& 0& 0& 1& 0& 0& 0& 0& 0& 0& 0& 0& 0\cr 
0& 0& 0& 0& 0& 0& 0& 0& 0& 1& 0& 0& 0& 0& 0& 0& 0& 0\cr 
0& 0& 0& 0& 0& 0& 0& 0& 0& 0& 1& 0& 0& 0& 0& 0& 0& 0\cr 
0& 0& 0& 0& 0& 0& 0& 0& 0& 0& 0& 1& 0& 0& 0& 0& 0& 0\cr 
0& 0& 0& 0& 0& 0& 0& 0& 0& 0& 0& 0& 1& 0& 0& 0& 0& 0\cr 
0& 0& 0& 0& 0& 0& 0& 0& 0& 0& 0& 0& 0& 1& 0& 0& 0& 0\cr 
0& 0& 0& 0& 0& 0& 0& 0& 0& 0& 0& 0& 0& 0& 1& 0& 0& 0\cr 
0& 0& 0& 0& 0& 0& 0& 0& 0& 0& 0& 0& 0& 0& 0& 1& 0& 0\cr 
0& 0& 0& 0& 0& 0& 0& 0& 0& 0& 0& 0& 0& 0& 0& 0& 1& 0\cr 
1& 2& 0& 2& 2& -3& -1& 1& 3& -2& 2& -1& 1& 3& 3& -1& 5& \zeta_1\cr 
-3& 2& 0& 2& -2& 1& 3& -3& -1& 2& 2& 3& 5& -1& -1& 3& 1& \zeta_2\cr
-3& -2& 0& -2& -2& 1& -5& 5& -1& 2& 2& 3& -3& -1& -1& 3& 1& \zeta_3\cr
1& -2& 0& -2& 2& 5& -1& 1& -5& -2& 2& -1& 1& 3&3&-1&-3&\zeta_4\cr
-3& -2& 0& 2& -2& 1& 3& -3& -1& 2& -2& 3& -3& -1& -1& -5& 1& \zeta_5\cr
1& -2& 0& 2& 2& -3& -1& 1& 3& -2& -2& -1& 1& 3&-5&-1&-3&\zeta_6\cr
1& 2& 0& -2& 2& -3& -1& 1& 3& -1& -1& -1& 1& -5& 3& -1& -3& \zeta_7\cr}
\right)
\end{array}
}
\]


\begin{thebibliography}{99} 
\bibitem{dkps}
M. R. Douglas, D. Kabat, P. Pouliot, S. Shenker,
``D-branes and Short Distances in String Theory,''
Nucl. Phys. {\bf B485} (1997) 85, {\tt hep-th/9608024}.
\bibitem{gk}
B. Greene, Y. Kanter, 
``Small Volumes in Compactified String Theory,''
Nucl. Phys. {\bf B497} (1997) 127, {\tt hep-th/9612181}. 
\bibitem{agm}
P. Aspinwall, B. Greene, D. Morrison, 
``Calabi-Yau Moduli Space, Mirror Manifolds and Space-time 
Topology Change in String Theory,''
Nucl. Phys. {\bf B416} (1994) 78, {\tt hep-th/9309097}.
\bibitem{wittenphases}
E. Witten, ``Phases of $N=2$ Theories in Two Dimensions,'' 
Nucl. Phys. {\bf B403} (1993) 159, {\tt hep-th/9301042}.
\bibitem{dgm}
M. Douglas, B. Greene, D. Morrison, 
``Orbifold Resolution by D-branes,''
Nucl. Phys. {\bf B506} (1997) 84, {\tt hep-th/9704151}.
\bibitem{wittenm}
E. Witten, 
``Phase Transitions in M Theory and F Theory,''
Nucl. Phys. {\bf B471} (1996) 195, {\tt hep-th/9603150}.
\bibitem{muto}
T. Muto, 
``D-branes on Orbifolds and Topology Change,''
Nucl.Phys. {\bf B521} (1998) 183, {\tt hep-th/9711090}. 
\bibitem{dm}
M. Douglas, G. Moore, 
``D-branes, Quivers, and ALE Instantons.''
{\tt hep-th/9603167}.
\bibitem{jm}
C. Johnson, R. Myers, 
``Aspects of Type IIB Theory on ALE Spaces,''
Phys. Rev. {\bf D55} (1997) 6382, {\tt hep-th/9610140}.
\bibitem{dg1}
M. Douglas, B. Greene, 
``Metrics on D-brane Orbifolds,''
Adv. Theor. Math. Phys. {\bf 1} (1998) 184, {\tt hep-th/9707214}. 
\bibitem{dko}
M. Douglas, A. Kato, H. Ooguri, Adv. Theor. Math. Phys. {\bf 1}
(1998) 237, {\tt hep-th/9708012}.
\bibitem{muto1}
T. Muto, ``Brane Configurations for Three-Dimensional Nonabelian
Orbifolds,'' {\tt hep-th/9905230}.
\bibitem{fhh1}
B. Feng, A. Hanany, Y. He, ``Z-D-brane Box Models and Nonchiral
Dihedral Quivers,'' {\tt hep-th/9909125}.
\bibitem{ot}
K. Oh, R. Tatar, ``Branes at Orbifolded Conifold Singularities 
and Supersymmetric Gauge Field Theories,'' JHEP {\bf 9910i} (1999)
0319, {\tt hep-th/9906012} 
\bibitem{celw}
M. Cvetic, L. Everett, P. Langacker, J. Wang,
``Blowing up the Four-Dimensional Z(3) Orientifold,''
JHEP {\bf 9904} (1999) 020, {\tt hep-th/9903051} 
\bibitem{fulton}
W. Fulton, {\tt Introduction to Toric Varieties}, Princeton University
Press, 1993.
\bibitem{cox}
D. Cox, {\tt alg-geom/9606016}.
\bibitem{gp}
E. Gimon, J. Polchinski, ``Consistency Conditions for Orientifolds 
and D Manifolds,'' Phys. Rev. {\bf D54} (1996) 1667, {\tt hep-th/9601038}.
\bibitem{gr1}
B. Greene, ``D-brane Topology Changing Transitions,'' Nucl.Phys. 
{\bf B525} (1998) 284, {\tt hep-th/9711124}.
\bibitem{mr}
S. Mukhopadhay, K. Ray, ``Conifolds from D-branes,''
Phys. Lett. {\bf B423} (1998) 247, {\tt hep-th/9711131}. 
\bibitem{fhh}
B. Feng, A. Hanany, Y. He, {\tt hep-th/0003085}.
\bibitem{bglp}
C. Beasley, B. Greene, C. Lazaroiu, M. Plesser, ``D3-branes on  
Partial Resolutions of Abelian Quotient Singularities of Calabi-Yau
Threefolds,'' Nucl. Phys. {\bf B553} (1999) 711, {\tt hep-th/9907186}.
\bibitem{pru}
J. Park, R. Rabadan, A. M. Uranga, hep-th/9907086, ``Orientifolding
the Conifold,'' Nucl.Phys. {\bf B570} (2000) 38, {\tt hep-th/9907086}. 
\bibitem{ur}
A. Uranga, ``Brane Configurtions for Branes at Conifolds,''
JHEP {\bf 9901} (1999) 022, {\tt hep-th/9811004} 
\bibitem{mp}
D. Morrison, R. Plesser, ``Nonspherical Horizons 1,''
Adv. Theor. Math. Phys. {\bf 3} (1999) 1, {\tt hep-th/9810201}.
\bibitem{ak}
A. Ahn, H. Kim, ``Branes at ${\bf C^4}/\Gamma$ Singularity from Toric
Geometry,'' JHEP {\bf 9904} (1999) 012, {\tt hep-th/9903181}.
\bibitem{mohri}
K. Mohri, ``D-branes and Quotient Singularities of Calabi-Yau
Fourfolds,'', Nucl. Phys. {\bf B521} (1998) 161, {\tt
hep-th/9707012}




\end{thebibliography}
\end{document}